\documentstyle[12pt]{article}
\newcommand{\nc}{\newcommand}
\nc{\al}{\alpha}
\nc{\ba}{\beta_\al}
\nc{\bb}{\beta_\beta}
\nc{\ga}{\g^\al}
\nc{\gb}{\g^\beta}
\nc{\db}{\pa_\beta}
\nc{\dtb}{\delta_\theta^\beta}
\nc{\dab}{{\delta_\al}^\beta}
\nc{\vmab}{V_{-\al}^\beta}
\nc{\vab}{V_\al^\beta}
\nc{\vib}{V_i^\beta}
\nc{\g}{\gamma}
\nc{\G}{\Gamma}
\nc{\D}{\Delta}
\nc{\paj}{P_{-\al}^j}
\nc{\la}{\lambda}
\nc{\La}{\Lambda}
\nc{\var}{\varphi}
\nc{\kvt}{\sqrt{t}}
\nc{\hn}{h^\vee}
\nc{\kn}{k^\vee}
\nc{\pa}{\partial}
\nc{\nn}{\nonumber \\ }
\nc{\hf}{\frac{1}{2}}         
\nc{\dz}{\frac{dz}{2\pi i}}
\nc{\fabc}{{f_{ab}}^c}
\nc{\bin}[2]{\left (\begin{array}{c} {#1}\\ {#2} \end{array}\right )}
\nc{\ben}{\begin{equation}}
\nc{\een}{\end{equation}}
\nc{\bea}{\begin{eqnarray}}
\nc{\eea}{\end{eqnarray}}
\nc{\bra}[1]{\langle {#1}|}
\nc{\ket}[1]{|{#1}\rangle}
\newcommand{\Z}{\mbox{$Z\hspace{-2mm}Z$}}
\nc{\C}{\mbox{\hspace{1.24mm}\rule{0.2mm}{2.5mm}\hspace{-2.7mm} C}}
\nc{\Nat}{\mbox{\hspace{.04mm}\rule{0.2mm}{2.8mm}\hspace{-1.5mm} N}}

\nc{\spa}{\hspace{1 cm},\hspace{1 cm}}
\nc{\vs}{\vspace}
\nc{\NP}[1]{Nucl.\ Phys.\ {\bf #1}}
\nc{\PL}[1]{Phys.\ Lett.\ {\bf #1}}
\nc{\CMP}[1]{Commun.\ Math.\ Phys.\ {\bf #1}}
\nc{\PR}[1]{Phys.\ Rev.\ {\bf #1}}
\nc{\PRL}[1]{Phys.\ Rev.\ Lett.\ {\bf #1}}
\nc{\PTP}[1]{Prog.\ Theor.\ Phys.\ {\bf #1}}
\nc{\PTPS}[1]{Prog.\ Theor.\ Phys.\ Suppl.\ {\bf #1}}
\nc{\MPL}[1]{Mod.\ Phys.\ Lett.\ {\bf #1}}
\nc{\IJMP}[1]{Int.\ Jour.\ Mod.\ Phys.\ {\bf #1}}
\nc{\IM}[1]{Invent.\ Math.\ {\bf #1}}
\nc{\SJNP}[1]{Sov. J. Nucl. Phys.\ {\bf #1}}

\begin{document}

\topmargin -5mm
\oddsidemargin 5mm

\begin{titlepage}
\setcounter{page}{0}
\begin{flushright}
July 1998
\end{flushright}

\vs{8mm}
\begin{center}
{\Large 3-point Functions in Conformal Field Theory}\\[.2cm]
{\Large with Affine Lie Group Symmetry}

\vs{8mm}
{\large J{\o}rgen Rasmussen}\footnote{e-mail address: 
jorgen@celfi.phys.univ-tours.fr}\\[.2cm]
{\em Laboratoire de Math\'{e}matiques et Physique Th\'{e}orique,}\\
{\em Universit\'{e} de Tours, Parc de Grandmont, F-37200 Tours, France}

\end{center}

\vs{8mm}
\centerline{{\bf{Abstract}}}
\noindent
In this paper we develop a general method for constructing 3-point functions
in conformal field theory with affine Lie group symmetry, continuing
our recent work on 2-point functions. The results are
provided in terms of triangular coordinates used in a wave function
description of vectors in highest weight modules. In this framework, 
complicated couplings translate into ordinary products of certain
elementary polynomials. The discussions
pertain to all simple Lie groups and arbitrary integrable representation. 
An interesting by-product is a general procedure for computing
tensor product coefficients, essentially by counting
integer solutions to certain inequalities.
As an illustration of the construction, we consider
in great detail the three cases $SL(3)$, $SL(4)$ and $SO(5)$.
\\[.4cm]

\end{titlepage}
\newpage
\renewcommand{\thefootnote}{\arabic{footnote}}
\setcounter{footnote}{0}

\section{Introduction}

The work presented in this paper constitutes a continuation of our
recent work on 2-point functions in conformal field theory with
affine Lie group symmetry or affine Lie algebra for short \cite{Ras1}. 
Our aim is to develop a systematic and general approach to the computation
of correlators. Thus, in \cite{Ras1} the program was initiated by 
considering 2-point functions for all Lie groups and arbitrary representations,
integrable or non-integrable. In this paper we undertake the study of 3-point
functions for integrable representations and present modest speculations
on a possible generalization to non-integrable representations.

Our method is based on a wave function formalism where vectors in 
certain representation spaces are translated into polynomials in so-called
triangular coordinates $x$. In this framework, complicated couplings of
highest weight representations are described by {\em ordinary} products
of the polynomials. We shall use the notion of elementary couplings
known from the study of tensor or Kronecker products of integrable
highest weight representations, see e.g. \cite{CCS,BMW,BKMW} and
references therein. This gives rise to a finite set of
elementary polynomials in terms of which our results may be expressed.
Certain redundancies, known as syzygies, appear in our formalism as
vanishing linear combinations of simple products of these elementary
polynomials. 

To the best of our knowledge, the only systematic approach to 3-point
functions in the literature is the work \cite{Rue}
by R\"uhl\footnote{We thank P. Furlan, A.Ch. Ganchev and V.B. Petkova 
for pointing out this work.}. However, his construction is designed
explicitly to treat integrable
representations in the case of $SL(N)$ only. In Section 4 we shall make
some comparative comments.

It should be stressed that our results are completely general wrt
the Lie algebra and the integrable representations, but pertain to
the {\em global} or horizontal
structure of the 3-point function. It is the dependence on the affine
coordinates $x$ that is determined in addition to the well known 
dependence on the conformal coordinates $z$. 
Thus, we do not
discuss the role played by the central extension (or level) $k$ of
the affine Lie algebra. Information on that is encoded in the 
structure constants which are yet to be determined. They are closely related
to the {\em physical} fusion coefficients. In the classical limit 
$k\rightarrow\infty$ these reduce to the ordinary {\em tensor} product 
coefficients. By construction, the latter denote the dimensionalities of the
spaces of chiral blocks that we obtain. Thus, in the process of constructing
the complete and minimal solution spaces for the 3-point functions we are also
determining the tensor product coefficients. This by-product is
an alternative to more conventional methods. 

Besides providing new and systematic results on 3-point functions,
our construction seems amenable of generalizing to non-integrable
representations. The indication is found in the very simple mathematical
structure of the correlators, where only linear combinations
of ordinary products of polynomials appear. As discussed in Section 5,
a generalization is currently under investigation.

The remaining part of this paper is organized as follows. In Section 2
we present some background material. In Section 3 we introduce the
wave function formalism and discuss the general structure of 3-point
functions in terms of elementary polynomials implementing the syzygies.
As building blocks, we introduce generalized anharmonic ratios. In Section
4 we discuss the cases $SL(3)$, $SO(5)$ and $SL(4)$ and present some
explicit results working out examples. Section 5 contains some concluding
speculations on possible generalizations, while various technical
results on $SO(5)$ and $SL(4)$ are listed in Appendix A.

\section{Notation}

Let {\bf g} be a simple Lie algebra of rank $r$.
{\bf h} is a Cartan subalgebra of {\bf g}. The set of (positive) roots
is denoted ($\Delta_+$) $\Delta$ and the simple roots are 
written $\al_i,\ i=1,...,r$. $\al^\vee = 2\al/\al^2$ is the root 
dual to $\al$. Using the triangular decomposition 
\ben
 \mbox{{\bf g}}=\mbox{{\bf g}}_-\oplus\mbox{{\bf h}}\oplus\mbox{{\bf g}}_+
\een
the raising and lowering operators are denoted $e_\al\in$ {\bf g}$_+$ and
$f_\al\in$ {\bf g}$_-$, respectively, with $\al\in\Delta_+$, and 
$h_i\in$ {\bf h} are the Cartan operators. 
In the Cartan-Weyl basis we have
\ben
 [h_i,e_\al]=(\al_i^\vee,\al)e_\al\spa [h_i,f_\al]=
  -(\al_i^\vee,\al)f_\al
\label{CW}
\een
and
\ben
 \left[e_\al,f_\al\right]=h_\al=G^{ij}(\al_i^\vee,\al^\vee)h_j
\een
where the metric $G_{ij}$ is related to the Cartan matrix $A_{ij}$ as
\ben
 A_{ij}=\al_i^\vee\cdot\al_j=(\al_i^\vee,\al_j)=
  G_{ij}\al_j^2/2
\een
The Dynkin labels $\La_k$ of the weight $\La$ are defined by
\ben
 \La=\La_k\La^{k}\spa \La_k=(\al_k^\vee,\Lambda)
\label{Dynkin}
\een
where $\left\{\La^{k}\right\}_{k=1,...,r}$ is the set of fundamental
weights satisfying
\ben
 (\al_i^\vee,\La^{k})=\delta_i^k
\een 
Elements in $\mbox{\bf g}_+$ may be parameterized using ``triangular 
coordinates" denoted by $x^\al$, one for each positive root, thus we 
write general Lie algebra elements in $\mbox{\bf g}_+$ as
\ben
 g_+(x)=x^\al e_\al \in \mbox{\bf g}_+
\label{gplus}
\een
We will understand ``properly" repeated root indices as in (\ref{gplus})
to be summed over the {\em positive} roots. Repeated Cartan indices as in
(\ref{Dynkin}) are also summed over. The corresponding group element is 
\ben
 G_+(x)=e^{g_+(x)}
\een
The matrix representation $C(x)$ of $g_+(x)$ in the adjoint 
representation is defined by
\ben
 C_a^b(x)=-x^\beta {f_{\beta a}}^b
\label{cadj}
\een

Now, a differential operator realization $\left\{\tilde{J}_a(x,\pa,\Lambda)
\right\}$ 
of the simple Lie algebra {\bf g} generated by $\left\{j_a\right\}$
may be defined by
\ben
 \bra{\La,x}j_a=\tilde{J_a}(x,\pa,\La)\bra{\La,x}
\een
where
\ben
 \bra{\La,x}=\bra{\La}G_+(x)
\een
The explicit solution is found to be \cite{PRY4}
\bea
\tilde{E}_\al(x,\pa)&=&\vab(x)\db\nn
\tilde{H}_i(x,\pa,\Lambda)&=&\vib(x)\db+\Lambda_i\nn
\tilde{F}_\al(x,\pa,\Lambda)&=&\vmab(x)\db+\paj(x)\Lambda_j
\label{defVP}
\eea
where 
\bea
 \vab(x)&=&\left[B(C(x))\right]_\al^\beta\nn
 \vib(x)&=&-\left[C(x)\right]_i^\beta \nn
 \vmab(x)&=&\left[e^{-C(x)}\right]_{-\al}^\g\left[B(-C(x))\right]_\g^\beta\nn
 \paj(x)&=&\left[e^{-C(x)}\right]_{-\al}^j 
\label{VPQ}
\eea
$B$ is the generating function for the Bernoulli numbers
\ben
  B(u)=\frac{u}{e^u-1}=\sum_{n\geq 0}\frac{B_n}{n!}u^n\nn
\label{Ber}
\een
whereas $\pa_\beta$ denotes partial differentiation wrt $x^\beta$.
Closely related to this differential operator realization is the equivalent
one $\left\{J_a(x,\pa,\Lambda)\right\}$ given by
\bea
 E_\al(x,\pa,\La)&=&-\tilde{F}_\al(x,\pa,\La)\nn
 F_\al(x,\pa,\La)&=&-\tilde{E}_\al(x,\pa,\La)\nn
 H_i(x,\pa,\La)&=&-\tilde{H}_i(x,\pa,\La)
\label{tilde}
\eea
The matrix functions (\ref{VPQ}) are defined in terms of universal
power series expansions, valid for any Lie algebra, but ones that truncate 
giving rise to finite polynomials of which the explicit forms depend on the
Lie algebra under consideration. Details on the truncations and the resulting
polynomials may be found in \cite{PRY4}.

\subsection{Affine Current Algebra}

Associated to a Lie algebra is an affine Lie algebra characterized by the
central extension
$k$, and associated to an affine Lie algebra is an affine current
algebra whose generators are conformal spin one fields and have amongst
themselves the operator product expansion
\ben
 J_a(z)J_b(w)=\frac{\kappa_{ab}k}{(z-w)^2}+\frac{\fabc J_c(w)}{z-w}
\label{JaJb}
\een
where regular terms have been omitted.
$\kappa_{ab}$ and $\fabc$ are the Cartan-Killing form and the 
structure coefficients, respectively, of the underlying Lie algebra.

It is convenient to collect the traditional multiplet of primary fields
in an affine current algebra (which generically is infinite for
non-integrable representations) in a generating function for that
\cite{FGPP,PRY1,PRY4}, namely the primary field 
$\phi_\La(w,x)$ which must satisfy
\bea
 J_a(z)\phi_\La(w,x)&=&\frac{-J_a(x,\pa,\La)}{z-w}\phi_\La(w,x)\nn
 T(z)\phi_\La(w,x)&=&\frac{\Delta(\phi_\La)}{(z-w)^2}\phi_\La(w,x)
  +\frac{1}{z-w}\pa\phi_\La(w,x)
\label{primdef}
\eea
Here $\left\{J_a(z)\right\}$ and $T(z)$ 
are the affine currents and the energy-momentum
tensor, respectively, whereas $\left\{J_a(x,\pa,\La)\right\}$ is the 
differential operator realization. $\D(\phi_\La)$ denotes the conformal
dimension of $\phi_\La$.
The explicit construction of primary fields for general simple Lie algebra
and arbitrary representation is provided in \cite{PRY4}, see also the 
discussion on wave functions below.

An affine transformation of a primary field is given by
\bea
 \delta_\epsilon\phi_\La(w,x)&=&\oint_w\frac{dz}{2\pi i}
  \epsilon^a(z)J_a(z)\phi_\La(w,x)\nn
 &=&\left\{\epsilon^{-\al}(w)V_\al^\beta(x)\pa_\beta
  +\epsilon^i(w)\left(V_i^\beta(x)
  \pa_\beta+\La_i\right)\right.\nn
  &+&\left.\epsilon^\al(w)\left(V_{-\al}^\beta(x)\pa_\beta+
  P_{-\al}^i(x)\La_i\right)\right\}\phi_\La(w,x)
\label{Ward}
\eea
and is parameterized by the $d$ ($d$ is the dimension of the
underlying Lie algebra) independent infinitesimal functions $\epsilon^a(z)$. 

\section{3-point Functions}

Let $W_3(z_1,z_2,z_3;x_1,x_2,x_3;\La_{(1)},\La_{(2)},\La_{(3)};k)$ 
denote a general 3-point function of the 3
primary fields $\phi_{\La_{(1)}}(z_1,x_1),\ \phi_{\La_{(2)}}(z_2,x_2)$ and
$\phi_{\La_{(3)}}(z_3,x_3)$.
Recall that $k$ is the central extension. The conformal Ward identities or
projective invariance allow us to determine completely the conformal property
($z$ dependence) of the 3-point function, whereby we may write
\bea
 &&W_3(z_1,z_2,z_3;x_1,x_2,x_3;\La_{(1)},\La_{(2)},\La_{(3)};k)\nn
 &=&C_{\La_{(1)}\La_{(2)}\La_{(3)}}(k)   
  (z_1-z_2)^{-\D_1-\D_2+\D_3}
  (z_2-z_3)^{-\D_2-\D_3+\D_1}
  (z_3-z_1)^{-\D_3-\D_1+\D_2}\nn
 &\cdot& W_n^{aff}(x_1,x_2,x_3;\La_{(1)},\La_{(2)},\La_{(3)})
\label{zdep}
\eea
where $\D_l=\D(\phi_{\La_{(l)}})$ is the conformal weight of the primary field
$\phi_{\La_{(l)}}$.
The affine Ward identity 
\bea
 0&=&\delta_\epsilon W_3(z_1,z_2,z_3;x_1,x_2,x_3;\La_{(1)},\La_{(2)},
  \La_{(3)};k)\nn
 &=&
 \langle\delta_\epsilon\phi_{\La_{(1)}}(z_1,x_1)\phi_{\La_{(2)}}(z_2,x_2)
  \phi_{\La_{(3)}}(z_3,x_3)
  \rangle+\langle\phi_{\La_{(1)}}(z_1,x_1)\delta_\epsilon
  \phi_{\La_{(2)}}(z_2,x_2)\phi_{\La_{(3)}}(z_3,x_3)\rangle\nn
 &+&\langle\phi_{\La_{(1)}}(z_1,x_1)\phi_{\La_{(2)}}(z_2,x_2)
  \delta_\epsilon\phi_{\La_{(3)}}(z_3,x_3)\rangle
\eea
may then be recast (using (\ref{Ward}))
into the following set of $d$ partial differential equations
\ben
 \left(\sum_{l=1}^3\tilde{J}_a(x_l,\pa,\La_{(l)})\right)
 W_3^{aff}(x_1,x_2,x_3;\La_{(1)},\La_{(2)},\La_{(3)})=0
\label{ddiff}
\een
It is easily verified that only the $2r$ equations for $a=\pm\al_i$
are independent. By induction, this simply
follows from the fact that $\{\tilde{J}_a\}$
is a differential operator realization of a Lie algebra.
It is the (finite dimensional)
solution spaces to the equations (\ref{ddiff}) for integrable 
representations that we shall discuss in the following. We shall use the
notation $W_3^{aff,m}$ to indicate the multiplicities. Thus, 
$m=1,...,N_{\La_{(1)}\La_{(2)}\La_{(3)}}$ where 
$N_{\La_{(1)}\La_{(2)}\La_{(3)}}$is the tensor product coefficient for
the coupling $(\La_{(1)},\La_{(2)},\La_{(3)})$. Note that we are only 
discussing the global ($k$ independent) behavior of the 
correlators, so $N_{\La_{(1)}\La_{(2)}\La_{(3)}}$ is also the 
dimension of the space of chiral blocks for the given coupling.
The non-trivial dependence on the central extension $k$ is encoded in
the coupling constants $C_{\La_{(1)}\La_{(2)}\La_{(3)}}(k)$. 
In general, also the
dimension of the space of {\em physical} chiral blocks will depend on
$k$ since it is given by the {\em fusion coefficient} 
$N_{\La_{(1)}\La_{(2)}\La_{(3)}}(k)$ (see e.g. \cite{BMW,BKMW} and
\cite{FGP}) satisfying
\ben
 N_{\La_{(1)}\La_{(2)}\La_{(3)}}(k)\leq N_{\La_{(1)}\La_{(2)}\La_{(3)}}
  (k=\infty)=N_{\La_{(1)}\La_{(2)}\La_{(3)}}
\een

\subsection{Wave Functions}

Here we shall discuss a wave function picture of the Kronecker or tensor
product of a set of integrable highest weight representations. The idea
\cite{PRY4,Ras1} is to translate states or vectors in the representation
spaces into polynomials in the triangular coordinates $x$, upon
which vectors in the tensor product modules are given by ordinary products
of polynomials. First we review
a few basic properties of conjugate weights and fundamental representations.

In every highest weight representation of highest weight $\La$ the weights 
are given
by $\la=\La-\sum\beta$ where $\sum\beta$ is a sum of positive roots or zero.
The depth of $\la$ is then defined as the height of $\sum\beta$. In a finite
dimensional irreducible highest weight module there exists a unique vector
(up to trivial renormalizations) of lowest weight characterized by
having maximal depth. The conjugate representation of such a representation
is a highest weight representation with highest weight $\La^+$ given by
minus the lowest weight of the original one, while in general all weights in 
the conjugate representation are given by minus the ones in the original
representation. The conjugate representation of a fundamental representation
(which is a finite dimensional irreducible highest weight representation
of highest weight a fundamental weight) is again a fundamental
representation. Due to the uniqueness of the conjugate weight
we shall write $\La^{i^+}=(\La^i)^+$. Many fundamental representations
(and therefore also many non-fundamental representations)
are self-conjugate, see e.g. \cite{Fuc}. In \cite{Ras1},
the natural generalization of the notion of conjugate weight to non-integrable
representations is discussed and the obvious result is
that the conjugate weight $\La^+$ to an arbitrary weight
$\La=\sum_{k=1}^r\La_k\La^k$, integrable or non-integrable, is given by
\ben
 \La^+=\sum_{k=1}^{r}\La_k^+\La^k=\sum_{k=1}^{r}\La_{k^+}\La^k
\label{conjla}
\een

Let $\La$ be an arbitrary integrable weight and let $\{ \ket{\la}\}^\mu$
denote a (not necessarily ortho-{\em normal})
basis in the highest weight module generated from the
highest weight vector $\ket{\La}$. A basis element may generically
be written
\ben
 \ket{\la}^\mu=f_{\al_{j_1}}...f_{\al_{j_{n(\mu)}}}\ket{\La}
\een
The corresponding wave function is defined by
\bea
 b(x,\La,\{j_l\}^\mu)&=&\bra{\La,x}\la\rangle^\mu\nn
 &=&\tilde{F}_{\al_{j_1}}(x,\pa,\La)...
  \tilde{F}_{\al_{j_{n(\mu)}}}(x,\pa,\La)1\nn
  &=&\left(V_{-\al_{j_1}}^\beta(x)\pa_\beta+P_{-\al_{j_1}}^{i_1}(x)\La_{i_1}
  \right)...\nn
   &\cdot&\left(V_{-\al_{j_{n(\mu)-1}}}^\beta(x)\pa_\beta+
    P_{-\al_{j_{n(\mu)-1}}}^{i_{n(\mu)-1}}(x)\La_{i_{n(\mu)-1}}\right)
  P_{-\al_{j_{n(\mu)}}}^{i_{n(\mu)}}(x)\La_{i_{n(\mu)}}
\label{wave}
\eea
Wave functions for elements in the tensor product of a set of 
integrable highest weight modules $(\La_{(1)},...,\La_{(n)})$
are then simply given by finite sums of products of wave functions 
\bea
 &&\sum_{\mu_1,...,\mu_n}C_{\mu_1...\mu_n}\ket{\la_{(1)}}^{\mu_1}\otimes...
 \otimes\ket{\la_{(n)}}^{\mu_n}\nn
 &\rightarrow&\bra{\La_{(1)},x_1}\otimes...\otimes\bra{\La_{(n)},x_n}
  \sum_{\mu_1,...,\mu_n}C_{\mu_1...\mu_n}\ket{\la_{(1)}}^{\mu_1}\otimes...
 \otimes\ket{\la_{(n)}}^{\mu_n}\nn
 &=&\sum_{\mu_1,...,\mu_n}C_{\mu_1...\mu_n}b(x_1,\La_{(1)},\left\{j_{l_1}
  \right\}^{\mu_1})...b(x_n,\La_{(n)},\left\{j_{l_n}
  \right\}^{\mu_n})
\eea
 
\subsection{Elementary Couplings and Polynomials}

When considering generating functions for tensor product coefficients
and branching rule multiplicities, fusion coefficients etc, elementary 
couplings are very useful, see \cite{CCS,BMW,BKMW} and references therein.
This is due to their basis properties: they are linearly independent,
all other couplings may be expressed in terms of them while they cannot
themselves be expressed in terms of other (elementary) couplings.
In the following we shall be interested mainly in 3-point couplings to
the scalar representation, though the notation introduced here allows for
discussion of general $n$-point couplings.

Let ${\cal L}_n^l$ denote the set of elementary $n$-point 
couplings involving the identity or zero weight exactly $n-l$ times.
The set of associated elementary polynomials (corresponding wave functions)
are then denoted ${\cal E}_n^l$. The total sets are denoted
\ben
 {\cal L}_n=\bigcup_{l=2}^n{\cal L}_n^l\spa 
  {\cal E}_n=\bigcup_{l=2}^n{\cal E}_n^l
\een
The element in ${\cal E}_n$ associated to the coupling $(\la_{(1)},...,
\la_{(n)})\in{\cal L}_n$ is denoted
\ben
 R^{\la_{(1)},...,\la_{(n)}}(x_1,...,x_n)\in{\cal E}_n
\een
The fundamental property of such an elementary polynomial 
is that for all Lie algebra generators  
\ben
 \left(\sum_{l=1}^n \tilde{J}_a(x_l,\pa,\la_{(l)})\right)
 R^{\la_{(1)},...,\la_{(n)}}(x_1,...,x_n)=0
\label{propR}
\een
For elements in ${\cal E}_n^2$, let us introduce the abbreviation
\ben
 R^i(x_j,x_k)=R^{\la_{(1)},...,\la_{(n)}}(x_1,...,x_n)
\ \  \mbox{if}\ \ \la_{(j)}
  =\La^i,\ \la_{(k)}=\La^{i^+}
\een

\subsubsection{Syzygies}

Even though the elements in ${\cal E}_n$ are linearly independent, there
may exist algebraic relations between them. They appear when a given coupling
may be represented non-uniquely in terms of elementary couplings, and are
conventionally denoted syzygies.
Correspondingly, we shall denote a non-trivial algebraic relation between 
elements of ${\cal E}_n$ a syzygy. Thus, in this wave function formalism
syzygies appear as vanishing linear combinations of ordinary products of 
elementary polynomials.

\subsection{Solutions in Terms of Generalized Anharmonic Ratios}

The main result in this paper is a prescription for generating the complete
and minimal set of solutions
to the following general ansatz for the affine
part of a 3-point function
\bea
 &&W_3^{aff}(x_1,x_2,x_3;\La_{(1)},\La_{(2)},\La_{(3)})\nn
 &=&\prod_{i=1}^{r}\left(
  \left(R^i(x_1,x_2)\right)^{p_i(1,2)}\left(R^i(x_2,x_3)\right)^{p_i(2,3)}
  \left(R^i(x_3,x_1)\right)^{p_i(3,1)}\right)\nn
 &\cdot&\prod_{(\la_{(1)},\la_{(2)},
 \la_{(3)})\in{\cal L}_3^3} \left(R^{\la_{(1)},\la_{(2)},\la_{(3)}}
  (x_1,x_2,x_3)
  \right)^{p_{\la_{(1)},\la_{(2)},\la_{(3)}}}
\label{ans}
\eea
where the exponents $p_i(1,2)$, $p_i(2,3)$, $p_i(3,1)$ and
$p_{\la_{(1)},\la_{(2)},\la_{(3)}}$ are 
functions of the weights $\La_{(1)},\ \La_{(2)}$ and $\La_{(3)}$.
The complete solution is given by a linear combination of such monomials
which, by construction, span the space of chiral blocks.

It is convenient to introduce
the following conjugation operation.
For any $r\times s$ matrix $M$ the conjugated $r\times s$ matrix 
$M^+$ is defined by interchanging the $i$'th and $j$'th rows when $i^+=j$,
that is when $\La^{i^+}=\La^j$ where $\{\La^i\}_{i=1}^r$ is the set of 
fundamental weights.
This is a straightforward generalization of the notion of the conjugate weight 
$\La^+$ to the weight $\La$ since in this respect
the weights are $r\times 1$ matrices.

Now, inserting the ansatz (\ref{ans}) in the affine Ward identities 
(\ref{ddiff}) leads to the
following linear equation system in the exponents 
\ben
 \left(\begin{array}{llll}
  I&0&I^+&M_1\\ I^+&I&0&M_2\\ 0&I^+&I&M_3 \end{array}\right) 
  \left(\begin{array}{l}p_1(1,2)\\ \vdots\\ p_r(1,2)\\ p_1(2,3)\\ \vdots \\
  p_r(2,3)\\ p_1(3,1)\\ \vdots\\ p_r(3,1)\\ p_{\la_{(1)},\la_{(2)},\la_{(3)}}
  \\ \vdots\\
 p_{\la_{(1)}',\la_{(2)}',\la_{(3)}'}\end{array}\right)
 =\left(\begin{array}{l}\La_{(1)}\\ \La_{(2)}\\ \La_{(3)} \end{array}\right)
\label{M}
\een
which may be recast into
\ben
 \left(\begin{array}{llcc}
  I&0&-I&M_2^+-M_3\\ 0&I&-I&-M_1^++M_2\\ 0&0&I+I^+&M_1-M_2^++M_3 
   \end{array}\right) 
  \left(\begin{array}{l}p_1(1,2)\\ \vdots\\ p_r(1,2)\\ p_1(2,3)\\ \vdots \\
  p_r(2,3)\\ p_1(3,1)\\ \vdots\\ p_r(3,1)\\ 
  p_{\la_{(1)},\la_{(2)},\la_{(3)}}\\ \vdots\\
 p_{\la_{(1)}',\la_{(2)}',\la_{(3)}'}\end{array}\right)
 =\left(\begin{array}{c}\La_{(2)}^+-\La_{(3)}\\ 
  -\La_{(1)}^++\La_{(2)}\\ \La_{(1)}-\La_{(2)}^++\La_{(3)} \end{array}\right)
\label{M2}
\een
Note that the matrices $M_1$, $M_2$ and 
$M_3$ thus defined are $r\times(\mbox{dim}\ ({\cal L}_3^3))$ matrices 
where any ordering of the elements of ${\cal L}_3^3$ may be chosen.
Since it is only certain representations in the $D_r$ series of classical
simple Lie algebras and most 
representations in the $A_r$ and $E_6$ series that are not self-conjugate,
the $B_r,C_r,E_7,E_8,F_4$ and $G_2$ series may be treated simultaneously
in which cases the set of linear equations become
\ben
 \left(\begin{array}{llcc}
  I&0&0&\hf\left(M_1+M_2-M_3\right)\\ 
  0&I&0&\hf\left(-M_1+M_2+M_3\right)\\ 
  0&0&I&\hf\left(M_1-M_2+M_3\right) 
   \end{array}\right) 
  \left(\begin{array}{l}p_1(1,2)\\ \vdots\\ p_r(1,2)\\ p_1(2,3)\\ \vdots \\
  p_r(2,3)\\ p_1(3,1)\\ \vdots\\ p_r(3,1)\\ 
  p_{\la_{(1)},\la_{(2)},\la_{(3)}}\\ \vdots\\
 p_{\la_{(1)}',\la_{(2)}',\la_{(3)}'}\end{array}\right)
 =\hf\left(\begin{array}{c}\La_{(1)}+\La_{(2)}-\La_{(3)}\\ 
  -\La_{(1)}+\La_{(2)}+\La_{(3)}\\ \La_{(1)}-\La_{(2)}+\La_{(3)} 
   \end{array}\right)
\label{selfc}
\een
The general solution to (\ref{M}) may be parameterized by $a_1,...,a_n$
where the number $n$ depends on the Lie algebra under consideration and
is given by the dimension of ${\cal L}_3$ subtracted the rank of the matrix
multiplying the exponent vector in (\ref{M}).
Correspondingly, there will be $n$ monomials $A_1,...,A_n$
in the elementary polynomials satisfying
\ben
 \left(\sum_{l=1}^3V_{-\al_i}^\beta(x_l)\pa_{x_l^\beta}\right)
  A_m=0\spa m=1,...,n
\label{anh}
\een
for all simple roots $\al_i$ (in fact for all roots).
These will be denoted generalized anharmonic ratios. The terminology is 
inspired by the well known anharmonic ratio in $SL(2)$ 
\ben
 z=\frac{(z_1-z_2)(z_3-z_4)}{(z_2-z_3)(z_4-z_1)}
\een
satisfying a condition similar to (\ref{anh}), namely
\bea
 \left(\sum_{l=1}^4V_{-\al}(z_l)\pa_{z_l}\right)z&=&0\nn
 V_{-\al}(z_l)&=&-(z_l)^2
\eea

In general, there are infinitely many solutions to the linear equation system
(\ref{M}). However, we are only interested in solutions pertaining to
{\em integrable} representations where all exponents in (\ref{ans}) 
must be non-negative integers, leaving only a {\em finite} set of
solutions. This number of solutions may in general exceed the
tensor product coefficient for the coupling in question. The reduction
is achieved by employing the syzygies. Indeed, following from those, the
generalized anharmonic ratios satisfy certain algebraic relations
which may be used in obtaining the final (reduced) set of independent 
3-point functions, see Section 4. 
Thus, an interesting by-product is a general procedure
for computing the tensor product coefficients, essentially by counting
integer solutions to the inequalities ensuring non-negative exponents, 
and with the complete sets of elementary couplings and associated syzygies 
as the only basic (group theoretical) input.

Our construction is illustrated in Section 4 where we consider the generically 
non-self-conjugate cases $SL(3)$ and $SL(4)$ (based on the Lie algebras
$A_2$ and $A_3$, respectively) and the self-conjugate case $SO(5)$ 
(based on the Lie algebra $B_2$). It should be stressed that the many 
explicit polynomials in Section 4 and Appendix A are only needed in order
to obtain explicit expressions for the syzygies. These again are only 
needed in establishing the algebraic relations satisfied by the 
generalized anharmonic ratios\footnote{The algebraic relations satisfied
by the anharmonic ratios depend on the normalizations of the elementary
polynomials. However, qua (\ref{wave})
these are fixed by the choice of basis for the
vectors in the highest weight module.}. Thus, it would be very convenient if
one could device a general procedure for determining the latter
relations or just the syzygies in our basis of elementary polynomials
without employing explicit knowledge of the polynomials. This is currently
under investigation.
Nevertheless, in order to perform the reductions, only the {\em form} is
needed of the algebraic relations satisfied by the generalized anharmonic
ratios. We shall comment further on this when discussing the case of
$SO(5)$ in Section 4.

\subsection{2-point Functions}

Here we review the result for the 2-point functions obtained in \cite{Ras1}.
The 2-point function of the primary fields $\phi_{\La_{(1)}}(z_1,x_1)$ and
$\phi_{\La_{(2)}}(z_2,x_2)$ in an affine current algebra is given by
\bea
 &&W_2(z_1,z_2;x_1,x_2;\La_{(1)},\La_{(2)};k)\nn
 &=&
  C_{\La_{(1)}\La_{(2)}}(k)
  \frac{\delta_{\D(\phi_{\La_{(1)}}),\D(\phi_{\La_{(2)}})}}{(z_1-z_2)^{
 \D(\phi_{\La_{(1)}})+\D(\phi_{\La_{(2)}})}}
 \prod_{i=1}^r\left(R^i(x_1,x_2)\right)^{p_i(\La_{(1)},\La_{(2)})}
\eea
where
\ben
 p_i(\La_{(1)},\La_{(2)})=\al_i^\vee\cdot\La_{(1)}
 =\al_{i^+}^\vee\cdot\La_{(2)}
\een
This result is valid for all representations, integrable or
non-integrable. It follows
immediately that the conjugate weight $\La^+$ to an arbitrary weight
$\La$, integrable or non-integrable, is given by (\ref{conjla}).

\section{Explicit Results for 3-point Functions}

In this section we shall illustrate our procedure by considering the 3
cases $SL(3)$, $SO(5)$ and $SL(4)$. Many technical details are referred
to Appendix A. The simplest case 
of $SL(2)$ is trivial since the loop projective invariance
fixes completely the $x$ dependence, to a form analogous to the $z$
dependence (\ref{zdep}). This is a well known result.
For higher groups and to the best of our knowledge, 
the work \cite{Rue} by R\"uhl is the only one discussing general
aspects of 3-point functions using triangular coordinates. However, his 
construction is designed explicitly to cover $SL(N)$ only and differs
considerably from ours, see below.

\subsection{Case of $SL(3)$}

The Lie algebra underlying affine $SL(3)$ current algebra is $A_2$
which has rank $r=2$ and one non-simple root, the highest root 
$\theta=\al_1+\al_2$. According to the general discussion above, we need
the explicit differential operator realization of the two lowering Chevalley 
generators which are easily found to be \cite{PRY4}
\bea
 \tilde{F}_{\al_1}(x,\pa,\La)&=&-(x^1)^2\pa_1+\left(\hf x^1x^2-x^\theta\right)
  \pa_2-\hf x^1\left(\hf x^1x^2+x^\theta\right)\pa_\theta+x^1\La_1\nn
 \tilde{F}_{\al_2}(x,\pa,\La)&=&\left(\hf x^1x^2+x^\theta\right)\pa_1
  -(x^2)^2\pa_2
  +\hf x^2\left(\hf x^1x^2-x^\theta\right)\pa_\theta+x^2\La_2
\eea
Here we have introduced the abbreviations $\pa_1=\pa_{\al_1}$ etc.
The elementary couplings are 
\bea
 {\cal L}_3^2&=&\left\{(\La^1,\La^2,0),(0,\La^1,\La^2),(\La^2,0,\La^1),\right.
  \nn
  &&\left. (\La^2,\La^1,0),(0,\La^2,\La^1),(\La^1,0,\La^2)\right\}\nn
 {\cal L}_3^3&=&\left\{(\La^1,\La^1,\La^1),(\La^2,\La^2,\La^2)\right\}
\eea
It is straightforward to work out explicit representations of the singlets. 
As an illustration,
\bea
 \ket{\mbox{singlet}(\La^1,\La^1,\La^1)}
 &=&\ket{1,0}\otimes\ket{-1,1}\otimes\ket{0,-1}-
  \ket{1,0}\otimes\ket{0,-1}\otimes\ket{-1,1}\nn
 &+&\ket{-1,1}\otimes\ket{0,-1}\otimes\ket{1,0}-
  \ket{-1,1}\otimes\ket{1,0}\otimes\ket{0,-1}\nn
 &+&\ket{0,-1}\otimes\ket{1,0}\otimes\ket{-1,1}-
  \ket{0,-1}\otimes\ket{-1,1}\otimes\ket{1,0}
\eea
is the singlet in the elementary coupling 
$(\La^1,\La^1,\La^1)\in{\cal L}_3^3$. Here the vector notation refers to the
elements in the fundamental module with highest weight $\La^1$ where 
a vector of weight $\La$ is denoted by its Dynkin labels
\ben
 \ket{\La}=\ket{\La_1,...,\La_r}
\een
The basis vectors in the two fundamental representation spaces are 
\ben
 \begin{array}{rlrl}
 \ket{1,0}&\hspace{4cm}&\ket{0,1}&\\
 \ket{-1,1}&=f_1\ket{1,0}&\ket{1,-1}&=f_2\ket{0,1}\\
 \ket{0,-1}&=f_2\ket{-1,1}&\ket{-1,0}&=f_1\ket{1,-1}
 \end{array}
\een
After having introduced the shorthand notation
\ben
 x_i^\pm=\hf x_i^1 x_i^2 \pm x_i^\theta
\een
the elementary polynomials may be written
\bea
 R^1(x_1,x_2)&=&x_1^++x_2^--x_1^1x_2^2\nn
 R^1(x_2,x_3)&=&x_2^++x_3^--x_2^1x_3^2\nn
 R^1(x_3,x_1)&=&x_3^++x_1^--x_3^1x_1^2\nn
 R^2(x_1,x_2)&=&x_1^-+x_2^+-x_1^2x_2^1\nn
 R^2(x_2,x_3)&=&x_2^-+x_3^+-x_2^2x_3^1\nn
 R^2(x_3,x_1)&=&x_3^-+x_1^+-x_3^2x_1^1\nn
 R^{1,1,1}(x_1,x_2,x_3)&=&x_2^1x_3^+-x_2^+x_3^1+x_1^1(x_2^+-x_3^+)+x_1^+
  (x_3^1-x_2^1)\nn
 R^{2,2,2}(x_1,x_2,x_3)&=&x_2^2x_3^--x_2^-x_3^2+x_1^2(x_2^--x_3^-)+x_1^-
  (x_3^2-x_2^2)
\eea
In this basis the general ansatz for the $x$ dependent part of the
3-point function becomes
\bea
 &&W_3^{aff}(x_1,x_2,x_3;\La_{(1)},\La_{(2)},\La_{(3)})\nn
 &=&\prod_{i=1,2}\left(
  \left(R^i(x_1,x_2)\right)^{p_i(1,2)}\left(R^i(x_2,x_3)\right)^{p_i(2,3)}
  \left(R^i(x_3,x_1)\right)^{p_i(3,1)}\right)\nn
 &\cdot&\left(R^{1,1,1}(x_1,x_2,x_3)\right)^{p_{1,1,1}}\left(
  R^{2,2,2}(x_1,x_2,x_3)\right)^{p_{2,2,2}}
\eea
It is easily verified that with the ordering $p_{1,1,1},p_{2,2,2}$
the 3 matrices $M_1$, $M_2$ and $M_3$ in (\ref{M}) are
all identical to the $2\times2$ unit matrix while the conjugate of any
$2\times s$ matrix is given by interchanging the two rows (since $(\La^1)^+
=\La^2$). Thus, the exponents are given by
\bea
 p_1(1,2)&=&\al_1^\vee\cdot\La_{(1)}-\frac{1}{3}\left(\al_1^\vee-
  \al_2^\vee\right)\cdot\left(\La_{(1)}+\La_{(2)}+\La_{(3)}\right)-(a_1+a_2)\nn
 p_2(1,2)&=&\al_1^\vee\cdot\La_{(2)}-\al_2^\vee\cdot\La_{(3)}
  -\frac{1}{3}\left(\al_1^\vee-
  \al_2^\vee\right)\cdot\left(\La_{(1)}+\La_{(2)}+\La_{(3)}\right)+a_1\nn
 p_1(2,3)&=&\al_2^\vee\cdot\La_{(3)}-(a_1+a_2)\nn
 p_2(2,3)&=&-\al_1^\vee\cdot\La_{(1)}+\al_2^\vee\cdot\La_{(2)}
  +\frac{1}{3}\left(\al_1^\vee-
  \al_2^\vee\right)\cdot\left(\La_{(1)}+\La_{(2)}+\La_{(3)}\right)+a_1\nn
 p_1(3,1)&=&\frac{1}{3}\al_1^\vee\cdot\left(\La_{(1)}-2\La_{(2)}
   +\La_{(3)}\right)
  +\frac{1}{3}\al_2^\vee\cdot\left(2\La_{(1)}-\La_{(2)}+2\La_{(3)}\right)
  -(a_1+a_2)\nn
 p_2(3,1)&=&a_1\nn
 p_{1,1,1}&=&\frac{1}{3}\left(\al_1^\vee-\al_2^\vee\right)\cdot\left(
   \La_{(1)}+\La_{(2)}+\La_{(3)}\right)+a_2\nn
 p_{2,2,2}&=&a_2
\label{psl3}
\eea
parameterized by $a_1$ and $a_2$, while
the generalized anharmonic ratios become
\bea
 A_1&=&\frac{R^2(x_1,x_2)R^2(x_2,x_3)R^2(x_3,x_1)}{R^1(x_1,x_2)R^1(x_2,x_3)
 R^1(x_3,x_1)}\nn
 A_2&=&\frac{R^{1,1,1}(x_1,x_2,x_3)R^{2,2,2}(x_1,x_2,x_3)}{
 R^1(x_1,x_2)R^1(x_2,x_3)R^1(x_3,x_1)}
\eea 
The single syzygy appears for the coupling 
$(\La^1+\La^2,\La^1+\La^2,\La^1+\La^2)$ and is worked out to be
\bea
 R^{1,1,1}(x_1,x_2,x_3)R^{2,2,2}(x_1,x_2,x_3)&=&R^1(x_1,x_2)R^1(x_2,x_3)
 R^1(x_3,x_1)\nn
 &+&R^2(x_1,x_2)R^2(x_2,x_3)R^2(x_3,x_1)
\eea
This means that the generalized anharmonic ratios are related as
\ben
 A_2=1+A_1
\een
It is recalled that $\La^1+\La^2=\theta$ is the adjoint representation.

Let us illustrate the procedure for obtaining the space of chiral blocks
by considering the coupling above where
\ben
 \La_{(1)}=\La_{(2)}=\La_{(3)}=\La^1+\La^2
\een
The condition that all powers (\ref{psl3}) should be non-negative
integers reduces to
\ben
 a_1,a_2,1-a_1-a_2\geq0
\een
with the obvious solution
\ben
 (a_1,a_2)\in\left\{(0,0),(0,1),(1,0)\right\}
\een
so the space of chiral blocks is generated by the functions
\ben
 \left\{1,A_1,A_2\right\}\rightarrow\left\{1,A_1\right\}
\een
The reduction is due to the syzygy. In conclusion, the space of chiral
blocks is 2 dimensional (in accordance with the well known result
that the corresponding tensor product coefficient is 2) and is spanned by
\bea
 W_3^{aff,1}&=&R^1(x_1,x_2)R^1(x_2,x_3)R^1(x_3,x_1)\nn
 W_3^{aff,2}&=&R^2(x_1,x_2)R^2(x_2,x_3)R^2(x_3,x_1)
\eea

\subsection{Case of $SO(5)$}

Here we shall consider the case $SO(5)$
where the affine current algebra is based on the
simple Lie algebra $B_2$. This Lie algebra has rank $r=2$ and
two non-simple positive roots denoted 
\ben
 \al_{11}=\al_1+\al_2\spa\theta=\al_1+2\al_2
\label{thetaB2}
\een
A discussion of $SO(5)$ may be found in \cite{Ras3} 
and the result presented there
relevant for our purpose is the following explicit expression for the 
differential operator realization of the two lowering Chevalley operators
\bea
 \tilde{F}_{\al_1}(x,\pa,\La)&=&-x^1x^1\pa_1+\left(\hf x^1x^2-x^{11}\right)
  \pa_2-\hf x^1\left(\hf x^1x^2+x^{11}\right)\pa_{11}\nn
 &+&\frac{1}{3}x^1x^2x^{11}\pa_\theta+x^1\La_1\nn
 \tilde{F}_{\al_2}(x,\pa,\La)&=&2\left(\hf x^1x^2+x^{11}\right)\pa_1-x^2x^2
  \pa_2+\left(\frac{1}{3}x^1x^2x^2-x^\theta\right)\pa_{11}\nn
 &-&x^2\left(\frac{1}{3}x^2x^{11}+
    x^\theta\right)\pa_\theta+x^2\La_2
\eea
Here we have introduced the abbreviations $x^{11}=x^{\al_{11}},\
\pa_1=\pa_{\al_1}$ etc.
The elementary couplings are 
\bea
 {\cal L}_3^2&=&\left\{(\La^1,\La^1,0),(\La^1,0,\La^1),(0,\La^1,\La^1),\right.
   \nn
  &&\left.(\La^2,\La^2,0),(0,\La^2,\La^2),(\La^2,0,\La^2)\right\}\nn
 {\cal L}_3^3&=&\left\{(\La^1,\La^2,\La^2),(\La^2,\La^1,\La^2),
   (\La^2,\La^2,\La^1),\right.\nn 
&&\left.(2\La^2,\La^1,\La^1),(\La^1,2\La^2,\La^1),(\La^1,\La^1,2\La^2)\right\}
\eea
Thus, the general ansatz for the affine part of the 3-point
function becomes 
\bea
 &&W_3^{aff}(x_1,x_2,x_3;\La_{(1)},\La_{(2)},\La_{(3)})\nn
 &=&\prod_{i=1,2}\left(
  \left(R^i(x_1,x_2)\right)^{p_i(1,2)}\left(R^i(x_2,x_3)\right)^{p_i(2,3)}
  \left(R^i(x_3,x_1)\right)^{p_i(3,1)}\right)\nn
 &\cdot&\left(R^{1,2,2}(x_1,x_2,x_3)\right)^{p_{1,2,2}}\left(
  R^{2,1,2}(x_1,x_2,x_3)\right)^{p_{2,1,2}}
  \left(R^{2,2,1}(x_1,x_2,x_3)\right)^{p_{2,2,1}}\nn
 &\cdot&\left(R^{4,1,1}(x_1,x_2,x_3)\right)^{p_{4,1,1}}\left(
  R^{1,4,1}(x_1,x_2,x_3)\right)^{p_{1,4,1}}
  \left(R^{1,1,4}(x_1,x_2,x_3)\right)^{p_{1,1,4}}
\eea
where $p_{1,2,2},p_{2,1,2},p_{2,2,1},p_{4,1,1},p_{1,4,1},p_{1,1,4}$ 
is a natural ordering of the exponents. Here we have
introduced the abbreviations $(1,2,2),(4,1,1)$ etc to denote the
elementary couplings $(\La^1,\La^2,\La^2),(2\La^2,\La^1,\La^1)$ etc.
Explicit expressions for the elementary polynomials are provided in
Appendix A. With this ansatz we find that
the three matrices $M_1$, $M_2$ and $M_3$ in (\ref{selfc}) are
\bea
 M_1&=&\left(\begin{array}{cccccc}
  1&0&0&0&1&1\\ 
  0&1&1&2&0&0\end{array}\right)\nn 
 M_2&=&\left(\begin{array}{cccccc}
  0&1&0&1&0&1\\ 
  1&0&1&0&2&0\end{array}\right)\nn
 M_3&=&\left(\begin{array}{cccccc}
  0&0&1&1&1&0\\ 
  1&1&0&0&0&2\end{array}\right)
\eea
The solution to (\ref{selfc}) in terms of the exponents may be written
\bea 
 p_1(1,2)&=&\hf\al_1^\vee\cdot\left(\La_{(1)}+\La_{(2)}-\La_{(3)}\right)
  -\hf\left(a_1+a_2-a_3+2a_6\right)\nn
 p_2(1,2)&=&\hf\al_2^\vee\cdot\left(\La_{(1)}+\La_{(2)}-\La_{(3)}\right)
  -\left(a_3+a_4+a_5-a_6\right)\nn
 p_1(2,3)&=&\hf\al_1^\vee\cdot\left(-\La_{(1)}+\La_{(2)}+\La_{(3)}\right)
  -\hf\left(-a_1+a_2+a_3+2a_4\right)\nn
 p_2(2,3)&=&\hf\al_2^\vee\cdot\left(-\La_{(1)}+\La_{(2)}+\La_{(3)}\right)
  -\left(a_1-a_4+a_5+a_6\right)\nn
 p_1(3,1)&=&\hf\al_1^\vee\cdot\left(\La_{(1)}-\La_{(2)}+\La_{(3)}\right)
  -\hf\left(a_1-a_2+a_3+2a_5\right)\nn
 p_2(3,1)&=&\hf\al_2^\vee\cdot\left(\La_{(1)}-\La_{(2)}+\La_{(3)}\right)
  -\left(a_2+a_4-a_5+a_6\right)\nn
 p_{1,2,2}&=&a_1\spa
 p_{2,1,2}=a_2\spa
 p_{2,2,1}=a_3\nn
 p_{4,1,1}&=&a_4\spa
 p_{1,4,1}=a_5\spa
 p_{1,1,4}=a_6
\eea
The generalized anharmonic ratios are readily seen to be
\bea
 A_1&=&\frac{(R^1(x_2,x_3))^\hf R^{1,2,2}(x_1,x_2,x_3)}{
  (R^1(x_1,x_2))^\hf R^2(x_2,x_3)(R^1(x_3,x_1))^\hf}\nn
 A_2&=&\frac{(R^1(x_3,x_1))^\hf R^{2,1,2}(x_1,x_2,x_3)}{
  (R^1(x_1,x_2))^\hf(R^1(x_2,x_3))^\hf R^2(x_3,x_1)}\nn
 A_3&=&\frac{(R^1(x_1,x_2))^\hf R^{2,2,1}(x_1,x_2,x_3)}{
  R^2(x_1,x_2)(R^1(x_2,x_3))^\hf(R^1(x_3,x_1))^\hf}\nn
 A_4&=&\frac{R^2(x_2,x_3)R^{4,1,1}(x_1,x_2,x_3)}{
  R^2(x_1,x_2)R^1(x_2,x_3)R^2(x_3,x_1)}\nn
 A_5&=&\frac{R^2(x_3,x_1)R^{1,4,1}(x_1,x_2,x_3)}{
  R^2(x_1,x_2)R^2(x_2,x_3)R^1(x_3,x_1)}\nn
 A_6&=&\frac{R^2(x_1,x_2)R^{1,1,4}(x_1,x_2,x_3)}{
  R^1(x_1,x_2)R^2(x_2,x_3)R^2(x_3,x_1)}
\eea
Due to the syzygies listed in Appendix A, the generalized
anharmonic ratios satisfy
the following set of non-trivial algebraic relations
\bea
 0&=&A_4+t_1 A_2A_3+u_1\nn
 0&=&A_5+t_2A_1A_3+u_2\nn
 0&=&A_6+t_3A_1A_2+u_3\nn
 0&=&A_1A_4+t_4A_2+u_4A_3\nn
 0&=&A_2A_5+t_5A_1+u_5A_3\nn
 0&=&A_3A_6+t_6A_1+u_6A_2\nn
 0&=&A_4A_5+t_7(A_3)^2+u_7\nn
 0&=&A_4A_6+t_8(A_2)^2+u_8\nn
 0&=&A_5A_6+t_9(A_1)^2+u_9
\label{so5anhrel}
\eea
The parameters $(u_\ell,t_\ell)$, $\ell=1,...,9$ are given in (\ref{so5ut})
in Appendix A.

Let us illustrate the procedure by considering the case 
\ben 
 \La_{(1)}=(0,2)\spa\La_{(2)}=(1,1)\spa\La_{(3)}=(1,1)
\een
One may show that the corresponding tensor product coefficient is 2.
Thus, we expect to find a 2 dimensional solution space 
for the chiral blocks. Now, the condition that all exponents should be
non-negative integers reduces to
\bea
 0&\leq&a_1,...,a_6\nn
 0&\leq&\hf\left(-a_1-a_2+a_3-2a_6\right)\nn
 0&\leq&1-a_3-a_4-a_5+a_6\nn
 0&\leq&1+\hf\left(a_1-a_2-a_3-2a_4\right)\nn
 0&\leq&-a_1+a_4-a_5-a_6\nn
 0&\leq&\hf\left(-a_1+a_2-a_3-2a_5\right)\nn
 0&\leq&1-a_2-a_4+a_5-a_6
\eea
with solution
\ben
 (a_1,a_2,a_3,a_4,a_5,a_6)\in\left\{(0,...,0),(0,1,1,0,0,0),(0,0,0,1,0,0)
  \right\}
\een
This means that the space of chiral blocks is generated by
\ben
 \left\{1,A_2A_3,A_4\right\}
\een
However, due to (\ref{so5anhrel}) this set may be reduced to
\ben
 \left\{1,A_4\right\}
\label{so5red}
\een
which agrees with the tensor product coefficient being 2.
In conclusion, the space of chiral blocks is 2 dimensional and is
spanned by
\bea
 W_3^{aff,1}&=&R^2(x_1,x_2)R^1(x_2,x_3)R^2(x_3,x_1)\nn
 W_3^{aff,2}&=&R^2(x_2,x_3)R^{4,1,1}(x_1,x_2,x_3)
\eea

A seemingly less trivial, yet simpler example is provided by the case
\ben
\La_{(1)}=(2,0)\spa\La_{(2)}=(0,2)\spa\La_{(3)}=(2,2)
\een
where we find that $a_l=0$, $l=1,...,6$ in order for the exponents to be
non-negative integers. Thus, the single chiral block is
\ben
 W_3^{aff}=\left(R^2(x_2,x_3)R^1(x_3,x_1)\right)^2
\een

Note that the reduction in (\ref{so5red}) may be inferred even without
knowing explicitly the parameters $(t_\ell,u_\ell)$ appearing in 
(\ref{so5anhrel})
(more generally $(s_\ell,t_\ell,u_\ell)$ etc, see Appendix A). All that is
required is the knowledge whether or not they are non-vanishing since 
the vanishing of a parameter alters the {\em form} of the syzygy.  
This general feature of our formalism
simplifies considerably the task of eliminating the redundancies due to
the existence of syzygies.
Nevertheless, in Appendix A we have provided the relevant parameters in
the cases $SO(5)$ and $SL(4)$.

\subsection{Case of $SL(4)$}

The Lie algebra underlying affine $SL(4)$ current algebra is $A_3$ which
has rank $r=3$ and three non-simple roots denoted 
\ben
 \al_{12}=\al_1+\al_2\spa\al_{23}=\al_2+\al_3\spa\theta=\al_1+\al_2+\al_3
\een
The relevant differential operators are worked out to be
\bea
 \tilde{F}_{\al_1}(x,\pa,\La)&=&-(x^1)^2\pa_1-\hf x^1\left(x^{12}+
  \hf x^1x^2\right)\pa_{12}+\left(\hf x^1x^2-x^{12}\right)\pa_2\nn
 &+&\left(-x^\theta+\hf x^1x^{23}+\frac{1}{12}x^1x^2x^3\right)\pa_{23}-
  \hf x^1\left(x^\theta+\frac{1}{2}x^1x^{23}+\frac{1}{6}x^{12}x^3
  \right)\pa_\theta\nn
 &+&x^1\La_1\nn
 \tilde{F}_{\al_2}(x,\pa,\La)&=&\left(x^{12}+\hf x^1x^2\right)\pa_1+\hf x^2
  \left(\hf x^1x^2-
  x^{12}\right)\pa_{12}-(x^2)^2\pa_2\nn
 &-&\hf x^2\left(\hf x^2x^3+x^{23}\right)\pa_{23}+\left(\hf x^2x^3-x^{23}
  \right)\pa_3+\frac{1}{6}x^2\left(x^1x^{23}-x^{12}x^3\right)\pa_\theta\nn
 &+&x^2\La_2\nn
 \tilde{F}_{\al_3}(x,\pa,\La)&=&\left(x^\theta+\hf x^{12}x^3-\frac{1}{12}
  x^1x^2x^3\right)
  \pa_{12}+\left(\hf x^2x^3+x^{23}\right)\pa_2-(x^3)^2\pa_3\nn
 &+&\hf x^3\left(\hf x^2x^3-x^{23}\right)\pa_{23}
  +\hf x^3\left(-x^\theta+\hf x^{12}x^3+\frac{1}{6}x^1x^{12}\right)\pa_\theta
  \nn
  &+&x^3\La_3
\eea
The elementary couplings are 
\bea
 {\cal L}_3^2&=&\left\{(\La^1,\La^3,0),(0,\La^1,\La^3),(\La^3,0,\La^1),\right.
  \nn
             &&(\La^2,\La^2,0),(0,\La^2,\La^2),(\La^2,0,\La^2),\nn
             &&\left.(\La^3,\La^1,0),(0,\La^3,\La^1),(\La^1,0,\La^3)\right\}\nn
 {\cal L}_3^3&=&\left\{(\La^1,\La^1,\La^2),(\La^1,\La^2,\La^1),
   (\La^2,\La^1,\La^1),\right.\nn
   &&(\La^2,\La^3,\La^3),(\La^3,\La^2,\La^3),(\La^3,\La^3,\La^2),\nn
   &&\left.(\La^1+\La^3,\La^2,\La^2),(\La^2,\La^1+\La^3,\La^2),
   (\La^2,\La^2,\La^1+\La^3)\right\}
\eea
Thus, the general ansatz for the affine part of the 3-point
function becomes 
\bea
 &&W_3^{aff}(x_1,x_2,x_3;\La_{(1)},\La_{(2)},\La_{(3)})\nn
 &=&\prod_{i=1}^3\left(
  \left(R^i(x_1,x_2)\right)^{p_i(1,2)}\left(R^i(x_2,x_3)\right)^{p_i(2,3)}
  \left(R^i(x_3,x_1)\right)^{p_i(3,1)}\right)\nn
 &\cdot&\left(R^{1,1,2}(x_1,x_2,x_3)\right)^{p_{1,1,2}}\left(
  R^{1,2,1}(x_1,x_2,x_3)\right)^{p_{1,2,1}}
  \left(R^{2,1,1}(x_1,x_2,x_3)\right)^{p_{2,1,1}}\nn
 &\cdot&\left(R^{2,3,3}(x_1,x_2,x_3)\right)^{p_{2,3,3}}\left(
  R^{3,2,3}(x_1,x_2,x_3)\right)^{p_{3,2,3}}
  \left(R^{3,3,2}(x_1,x_2,x_3)\right)^{p_{3,3,2}}\nn
 &\cdot&\left(R^{4,2,2}(x_1,x_2,x_3)\right)^{p_{4,2,2}}\left(
  R^{2,4,2}(x_1,x_2,x_3)\right)^{p_{2,4,2}}
  \left(R^{2,2,4}(x_1,x_2,x_3)\right)^{p_{2,2,4}} 
\eea
where $p_{1,1,2},p_{1,2,1},p_{2,1,1},p_{2,3,3},p_{3,2,3},p_{3,3,2},
p_{4,2,2},p_{2,4,2},p_{2,2,4}$
is a natural ordering of the exponents. Here we have introduced the 
abbreviations $(1,1,2),(2,4,2)$ etc to denote
the elementary couplings $(\La^1,\La^1,\La^2),(\La^2,\La^1+\La^3,\La^2)$ etc.
Explicit expressions for the elementary polynomials are provided
in Appendix A. We find the following solution to the linear equation
system (\ref{M})
\bea
 p_1(1,2)&=&\hf\al_1^\vee\cdot\left(\La_{(1)}-\La_{(2)}-\La_{(3)}\right)
  +\hf\al_3^\vee\cdot\left(\La_{(1)}+\La_{(2)}+\La_{(3)}\right)\nn
 &-&(a_1-a_3+a_4+a_5+a_6+a_7)\nn
 p_2(1,2)&=&\hf\al_2^\vee\cdot\left(\La_{(1)}+\La_{(2)}-\La_{(3)}\right)
  +\frac{1}{4}\left(\al_1^\vee-\al_3^\vee\right)\cdot\left(\La_{(1)}
  +\La_{(2)}+\La_{(3)}\right)\nn
 &-&(a_2+a_3-a_6+a_9)\nn
 p_3(1,2)&=&\hf\al_1^\vee\cdot\left(-\La_{(1)}+\La_{(2)}-\La_{(3)}\right)
  +\hf\al_3^\vee\cdot\left(\La_{(1)}+\La_{(2)}-\La_{(3)}\right)\nn
 &-&(-a_1-a_2+a_6+a_8-a_9)\nn  
 p_1(2,3)&=&\al_3^\vee\cdot\La_{(3)}
  -(a_1+a_4+a_5+a_9)\nn
 p_2(2,3)&=&\hf\al_2^\vee\cdot\left(-\La_{(1)}+\La_{(2)}+\La_{(3)}\right)
  -\frac{1}{4}\left(\al_1^\vee-\al_3^\vee\right)\cdot\left(\La_{(1)}
  +\La_{(2)}+\La_{(3)}\right)\nn
 &-&(-a_3+a_5+a_6+a_7)\nn
 p_3(2,3)&=&\hf\al_1^\vee\cdot\left(-\La_{(1)}+\La_{(2)}+\La_{(3)}\right)
  +\hf\al_3^\vee\cdot\left(-\La_{(1)}+\La_{(2)}-\La_{(3)}\right)\nn
 &-&(-a_1+a_3-a_5-a_7+a_8)\nn
 p_1(3,1)&=&\hf\left(\al_1^\vee+\al_3^\vee\right)
  \cdot\left(\La_{(1)}-\La_{(2)}+\La_{(3)}\right)\nn
 &-&(a_1+a_2+a_5+a_7-a_8+a_9)\nn
 p_2(3,1)&=&\hf\al_2^\vee\cdot\left(\La_{(1)}-\La_{(2)}+\La_{(3)}\right)
  -\frac{1}{4}\left(\al_1^\vee-\al_3^\vee\right)\cdot\left(\La_{(1)}
  +\La_{(2)}+\La_{(3)}\right)\nn
 &-&(-a_2+a_4+a_6+a_8)\nn
 p_3(3,1)&=&a_1\nn
 p_{1,1,2}&=&\hf\left(\al_1^\vee-\al_3^\vee\right)\cdot\left(\La_{(1)}
  +\La_{(2)}+\La_{(3)}\right)\nn
 &-&(a_2+a_3-a_4-a_5-a_6)\nn
 p_{1,2,1}&=&a_2\spa
 p_{2,1,1}=a_3\nn
 p_{2,3,3}&=&a_4\spa
 p_{3,2,3}=a_5\spa
 p_{3,3,2}=a_6\nn
 p_{4,2,2}&=&a_7\spa
 p_{2,4,2}=a_8\spa
 p_{2,2,4}=a_9
\eea 
The generalized anharmonic ratios are readily seen to be
\bea
 A_1&=&\frac{R^3(x_1,x_2)R^3(x_2,x_3)R^3(x_3,x_1)}{
  R^1(x_1,x_2)R^1(x_2,x_3)R^1(x_3,x_1)}\nn
 A_2&=&\frac{R^3(x_1,x_2)R^2(x_3,x_1)R^{1,2,1}(x_1,x_2,x_3)}{
  R^2(x_1,x_2)R^1(x_3,x_1)R^{1,1,2}(x_1,x_2,x_3)}\nn
 A_3&=&\frac{R^1(x_1,x_2)R^2(x_2,x_3)R^{2,1,1}(x_1,x_2,x_3)}{
  R^2(x_1,x_2)R^3(x_2,x_3)R^{1,1,2}(x_1,x_2,x_3)}\nn
 A_4&=&\frac{R^{1,1,2}(x_1,x_2,x_3)R^{2,3,3}(x_1,x_2,x_3)}{
  R^1(x_1,x_2)R^1(x_2,x_3)R^2(x_3,x_1)}\nn
 A_5&=&\frac{R^3(x_2,x_3)R^{1,1,2}(x_1,x_2,x_3)R^{3,2,3}(x_1,x_2,x_3)}{
  R^1(x_1,x_2)R^1(x_2,x_3)R^2(x_2,x_3)R^1(x_3,x_1)}\nn
 A_6&=&\frac{R^2(x_1,x_2)R^{1,1,2}(x_1,x_2,x_3)R^{3,3,2}(x_1,x_2,x_3)}{
  R^1(x_1,x_2)R^3(x_1,x_2)R^2(x_2,x_3)R^2(x_3,x_1)}\nn
 A_7&=&\frac{R^3(x_2,x_3)R^{4,2,2}(x_1,x_2,x_3)}{
  R^1(x_1,x_2)R^2(x_2,x_3)R^1(x_3,x_1)}\nn
 A_8&=&\frac{R^1(x_3,x_1)R^{2,4,2}(x_1,x_2,x_3)}{
  R^3(x_1,x_2)R^3(x_2,x_3)R^2(x_3,x_1)}\nn
 A_9&=&\frac{R^3(x_1,x_2)R^{2,2,4}(x_1,x_2,x_3)}{
  R^2(x_1,x_2)R^1(x_2,x_3)R^1(x_3,x_1)}
\eea
Due to the syzygies listed in Appendix A, the generalized anharmonic ratios
satisfy the following set of non-trivial algebraic relations
\bea
 0&=&A_7A_8+t_1A_6+u_1\nn
 0&=&A_8A_9+t_2A_3A_4+u_2\nn
 0&=&A_7A_9+t_3A_2A_5+u_3A_1\nn
 0&=&A_4A_7+t_4A_1A_6+u_4A_5\nn
 0&=&A_5A_8+t_5A_4+u_5A_6\nn
 0&=&A_6A_9+t_6A_5+u_6A_4\nn
 0&=&A_9+t_7A_1A_3+u_7A_2\nn
 0&=&A_2A_8+t_8+u_8A_3\nn
 0&=&A_3A_7+t_9A_2+u_9\nn
 0&=&A_7+t_{10}A_5+u_{10}A_1\nn
 0&=&A_9+t_{11}A_2A_4+u_{11}A_1\nn
 0&=&A_8+t_{12}A_3A_6+u_{12}\nn
 0&=&A_1A_8+t_{13}A_4+u_{13}\nn
 0&=&A_9+t_{14}A_3A_5+u_{14}\nn
 0&=&A_7+t_{15}A_2A_6+u_{15}
\label{sl4anhrel}
\eea
The parameters $(u_\ell,t_\ell)$, $\ell=1,...,15$ are given in (\ref{sl4ut})
in Appendix A. 

Let us illustrate the procedure by considering the case 
\ben 
 \La_{(1)}=(1,2,1)\spa\La_{(2)}=(1,2,1)\spa\La_{(3)}=(1,2,1)
\een
One may show that the corresponding tensor product coefficient is 5, e.g.
by counting independent Berenstein-Zelevinsky triangles \cite{BZ} as
discussed in \cite{BKMW}.
Thus, we expect to find a 5 dimensional solution space 
for the chiral blocks. Now, the condition that all exponents should be
non-negative integers reduces to
\bea
 0&\leq&a_1,...,a_9\nn
 0&\leq&1-a_1+a_3-a_4-a_5-a_6-a_7\nn
 0&\leq&1-a_2-a_3+a_6-a_9\nn
 0&\leq&a_1+a_2-a_6-a_8+a_9\nn
 0&\leq&1-a_1-a_4-a_5-a_9\nn
 0&\leq&1+a_3-a_5-a_6-a_7\nn
 0&\leq&a_1-a_3+a_5+a_7-a_8\nn
 0&\leq&1-a_1-a_2-a_5-a_7+a_8-a_9\nn
 0&\leq&1+a_2-a_4-a_6-a_8\nn
 0&\leq&-a_2-a_3+a_4+a_5+a_6
\eea
with solution
\bea
 &&(a_1,a_2,a_3,a_4,a_5,a_6)\nn
 &\in&\left\{(0,...,0),(1,0,...,0),(0,0,0,1,0,...,0),
  (0,0,0,0,1,0,0,0,0)\right.\nn
 &&(0,...,0,1,0,0),(0,...,0,1),(1,0,...,0,1,0),(0,1,0,1,0,...,0)\nn
 &&\left.(0,1,0,0,0,1,0,0,0),(0,0,1,0,1,0,...,0),(0,...,0,1,0,0,1)\right\}
\eea
This means that the space of chiral blocks is generated by
\ben
 \left\{1,A_1,A_4,A_5,A_7,A_9,A_1A_8,A_2A_4,A_2A_6,A_3A_5,A_6A_9\right\}
\een
However, due to (\ref{sl4anhrel}) this set may be reduced to
\ben
 \left\{1,A_1,A_4,A_7,A_9\right\}
\een
which agrees with the tensor product coefficient being 5.
In conclusion, the space of chiral blocks is 5 dimensional and is
spanned by
\bea
 W_3^{aff,1}&=&R^1(x_1,x_2)R^2(x_1,x_2)R^1(x_2,x_3)R^2(x_2,x_3)R^1(x_3,x_1)
  R^2(x_3,x_1)\nn
 W_3^{aff,2}&=&R^2(x_1,x_2)R^3(x_1,x_2)R^2(x_2,x_3)R^3(x_2,x_3)R^2(x_3,x_1)
  R^3(x_3,x_1)\nn
 W_3^{aff,3}&=&R^2(x_1,x_2)R^2(x_2,x_3)R^1(x_3,x_1)R^{1,1,2}(x_1,x_2,x_3)
  R^{2,2,3}(x_1,x_2,x_3)\nn
 W_3^{aff,4}&=&R^2(x_1,x_2)R^1(x_2,x_3)R^3(x_2,x_3)R^2(x_3,x_1)
  R^{4,2,2}(x_1,x_2,x_3)\nn
 W_3^{aff,5}&=&R^1(x_1,x_2)R^3(x_1,x_2)R^2(x_2,x_3)R^2(x_3,x_1)
  R^{2,2,4}(x_1,x_2,x_3)
\eea

{}From a purely algebraic point of view, it is sufficient to consider $A_1$, 
$A_4$ and $A_7$ as independent building block functions since the remaining
6 anharmonic ratios in addition to the constant function may be expressed
in terms of these 3 by employing (\ref{sl4anhrel}). As an illustration we
find
\ben
 A_6=-\frac{1}{t_4}\frac{A_4A_7}{A_1}+\frac{u_4}{t_4t_{10}}\left(
  u_{10}+\frac{A_7}{A_1}\right)
\een
This observation indicates that an alternative but related procedure
exists for producing 3-point functions in the case of $SL(4)$. Even though
it is based on 3 building block functions (anharmonic ratios) only, it seems
less transparent. In the work \cite{Rue} by R\"uhl a procedure of that
kind is discussed. It is designed explicitly to cover $SL(N)$ and he finds
that the number of building block functions (in addition to equivalences to
the elementary 2-point polynomials $R^i(x_j,x_k)$) is equal to the number
of solutions to the constraint system
\ben
 p_1,p_2,p_3\in\Z\spa1\leq p_1,p_2,p_3\leq N\spa p_1+p_2+p_3=N
\een
Indeed, in the case of $SL(4)$ this number is 3. Let us emphasize
that \cite{Rue} only covers 2-point and 3-point functions for $SL(N)$
and that examples of 3-point functions are only worked out for $SL(3)$.

\section{Conclusion}

We shall refrain from repeating the results presented in this paper
pertaining to {\em integrable} representation. Instead, we shall
comment on possible generalizations.

Compared to the case of integrable representations, very little is known about
non-integrable representations. Nevertheless, degenerate \cite{KK} and
in particular admissible representations \cite{KW}
appear naturally in certain
physical applications. An important example is provided by Hamiltonian
reduction where generalized ($W$ extended) minimal models in CFT are obtained
by constraining affine current algebras for admissible representations. 
The entire regime of the Coulomb gas picture is obtained by considering 
the broader class of degenerate representations. Thus, it is of great
interest to discuss generalizations from integrable to non-integrable 
representations. Hamiltonian reduction at the level of correlators has
been discussed in \cite{FGPP,PRY2,FGPP2}

We believe that the framework of wave functions employed in this paper (and in
\cite{PRY4,Ras1}) is suitable for considering this troublesome generalization.
Our belief is mainly based on the fundamental property of the formalism
that all questions for integrable representations
are translated into the simple language of
finite polynomials and derivatives thereof. A natural proposal is to allow
for finite products of monomials in the elementary polynomials raised 
to {\em non-integer} powers given by the generically non-integer Dynkin 
labels. From the mathematical point of view, 
such a situation is equally well under control, though some
non-trivial questions arise, e.g. 
the dimensionality of the solution space for the chiral blocks.
In the case of 2-point functions the generalization is straightforward and
discussed in \cite{Ras1}, whereas the case of 3-point functions and eventually 
higher point functions are currently under investigation. Also generalizations
to supergroups are currently being investigated where the analogous wave 
function picture is described in \cite{Ras2}. 

Furthermore, we hope to
come back elsewhere with a discussion on higher point functions for
integrable representations, in particular 4-point functions. Hamiltonian
reduction will then yield valuable new results in Toda theory.\\[.4cm]
{\bf Acknowledgment}\\[.2cm]
The author thanks Jens Schnittger for fruitful discussions and 
Pierre Mathieu for fruitful correspondence
and for pointing out reference \cite{CCS}. He  
gratefully acknowledges the financial support from the Danish 
Natural Science Research Council, contract no. 9700517.

\appendix
\section{Examples of Elementary Polynomials and Their Associated Syzygies}

\subsection{Case of $SO(5)$}

We find the following 12 elementary polynomials
\bea
 R^{\La^1,\La^1,0}&=&-2x_2^\theta x_2^1-2(x_2^{11})^2
  +\frac{1}{6}(x_2^1x_2^2)^2-x_1^1\left(-2x_2^\theta+2x_2^{11}x_2^2
  +\frac{2}{3}x_2^1(x_2^2)^2\right)\nn
 &+&\left(2x_1^{11}+x_1^1x_1^2\right)\left(2x_2^{11}+x_2^1x_2^2\right)
  -\left(-2x_1^\theta+2x_1^{11}x_1^2+\frac{2}{3}x_1^1(x_1^2)^2\right)x_2^1\nn
 &-&2x_1^\theta x_1^1-2(x_1^{11})^2+\frac{1}{6}(x_1^1x_1^2)^2\nn
 R^{\La^2,\La^2,0}&=&x_2^\theta+\frac{1}{6}x_2^1(x_2^2)^2-
  x_1^2\left(\hf x_2^1x_2^2-x_2^{11}\right)
  +\left(\hf x_1^1x_1^2-x_1^{11}\right)x_2^2-x_1^\theta-\frac{1}{6}
  x_1^1(x_1^2)^2\nn
 R^{\La^1,\La^2,\La^2}&=&\left(x_2^1x_2^2-2x_2^{11}\right)
  \left(x_3^\theta+\frac{1}{6}x_3^1(x_3^2)^2\right)
  -\left(x_2^\theta+\frac{1}{6}x_2^1(x_2^2)^2\right)\left(x_3^1x_3^2
  -2x_3^{11}\right)\nn
 &+&2x_1^1\left(x_2^\theta+\frac{1}{6}x_2^1(x_2^2)^2\right)
   x_3^2-2x_1^1x_2^2\left(x_3^\theta+\frac{1}{6}x_3^1(x_3^2)^2\right)\nn
 &+&\left(-x_1^\theta+x_1^{11}x_1^2+\frac{1}{3}x_1^1(x_1^2)^2\right)
  \left(x_2^1x_2^2-2x_2^{11}\right)\nn
 &-&\left(-x_1^\theta+x_1^{11}x_1^2+\frac{1}{3}
   x_1^1(x_1^2)^2\right)\left(x_3^1x_3^2-2x_3^{11}\right)\nn
 &+&\left(-2x_1^\theta x_1^1-2(x_1^{11})^2+\frac{1}{6}
  (x_1^1x_1^2)^2\right)x_3^2\nn
 &-&\left(-2x_1^\theta x_1^1-2x_1^{11}x_1^{11}
  +\frac{1}{6}(x_1^1x_1^2)^2\right)x_2^2\nn
 &+&\left(2x_1^{11}+
  x_1^1x_1^2\right)\left(x_3^\theta+\frac{1}{6}x_3^1(x_3^2)^2\right)
  -\left(2x_1^{11}+x_1^1x_1^2\right)\left(x_2^\theta+\frac{1}{6}x_2^1
  (x_2^2)^2\right)\nn
 &+&\left(x_1^{11}+\frac{1}{3}x_1^1(x_1^2)^2\right)x_2^2
   \left(x_3^1x_3^2-2x_3^{11}\right)
   -\left(2x_1^{11}+x_1^1x_1^2\right)\left(\hf x_2^1x_2^2-x_2^{11}\right)
  x_3^2\nn 
 R^{2\La^2,\La^1,\La^1}&=&2\left(-2x_2^\theta+2x_2^{11}(x_2^2)^2
  +\frac{2}{3}x_2^1(x_2^2)^2\right)\left(-2x_3^\theta x_3^1-2(x_3^{11})^2
  +\frac{1}{6}(x_3^1x_3^2)^2\right)\nn
 &-&2\left(-2x_2^\theta x_2^1-2(x_2^{11})^2+\frac{1}{6}(x_2^1x_2^2)^2
  \right)\left(-2x_3^\theta+2x_3^{11}x_3^2+\frac{2}{3}x_3^1(x_3^2)^2
  \right)\nn
 &+&4x_1^2\left(-2x_2^\theta x_2^1-2(x_2^{11})^2+\frac{1}{6}(x_2^1
  x_2^2)^2\right)\left(2x_3^{11}+x_3^1x_3^2\right)\nn
 &-&4x_1^2\left(2x_2^{11}+x_2^1x_2^2\right)\left(-2x_3^\theta x_3^1
  -2(x_3^{11})^2+\frac{1}{6}(x_3^1x_3^2)^2\right)\nn
 &+&2\left(x_1^1x_1^2-2x_1^{11}\right)\left(2x_2^{11}+x_2^1x_2^2\right)
  \left(-2x_3^\theta+2x_3^{11}x_3^2+\frac{2}{3}x_3^1(x_3^2)^2\right)\nn
 &-&2\left(x_1^1x_1^2-2x_1^{11}\right)
  \left(-2x_2^\theta+2x_2^{11}x_2^2+\frac{2}{3}x_2^1(x_2^2)^2\right)
  \left(2x_3^{11}+x_3^1x_3^2\right)\nn
 &+&4(x_1^2)^2x_2^1\left(-2x_3^\theta x_3^1-2(x_3^{11})^2
  +\frac{1}{6}(x_3^1x_3^2)^2\right)\nn
 &-&4x_1^2x_1^2\left(-2x_2^\theta x_2^1-2(x_2^{11})^2
  +\frac{1}{6}(x_2^1x_2^2)^2\right)x_3^1\nn
 &+&2\left(4x_1^2x_1^\theta+\frac{2}{3}x_1^1(x_1^2)^3\right)\left(
  x_2^1\left(2x_3^{11}+x_3^1x_3^2\right)-\left(2x_2^{11}+x_2^1x_2^2\right)
  x_3^1\right)\nn
 &+&\left(-4x_1^1x_1^2x_1^{11}+4(x_1^{11})^2+(x_1^1x_1^2)^2\right)\nn
 &\cdot&\left(2x_2^\theta-2x_2^{11}x_2^2-\frac{2}{3}x_2^1(x_2^2)^2
  -2x_3^\theta+2x_3^{11}x_3^2+\frac{2}{3}x_3^1(x_3^2)^2\right)\nn
 &+&\left(4x_1^\theta x_1^1x_1^2-8x_1^\theta x_1^{11}-\frac{4}{3}x_1^1
  (x_1^2)^2x_1^{11}+\frac{2}{3}(x_1^1)^2(x_1^2)^3\right)\nn
 &\cdot&\left(2x_2^{11}+x_2^1x_2^2-2x_3^{11}-x_3^1x_3^2\right)\nn
 &+&\left(8(x_1^\theta)^2-\frac{4}{3}x_1^1(x_1^2)^3x_1^{11}
  +\frac{8}{3}x_1^1(x_1^2)^2x_1^\theta-\frac{4}{9}(x_1^1)^2(x_1^2)^4\right)
  \left(x_3^1-x_2^1\right)\nn
 &+&4x_1^2\left(x_1^1x_1^2-2x_1^{11}\right)\nn
 &\cdot& \left(-2x_2^\theta x_2^1-2(x_2^{11})^2+\frac{1}{6}(x_2^1
  x_2^2)^2+2x_3^\theta x_3^1-2(x_3^{11})^2+\frac{1}{6}(x_3^1x_3^2)^2
  \right)\nn
 &+&2\left(2x_1^\theta-2x_1^{11}x_1^2+\frac{4}{3}x_1^1(x_1^2)^2\right)\nn
 &\cdot&\left\{2x_2^\theta x_2^1+2(x_2^{11})^2-\frac{1}{6}(x_2^1x_2^2)^2
  -2x_3^\theta x_3^1-2(x_3^{11})^2+\frac{1}{6}(x_3^1x_3^2)^2\right.\nn
 &+&\left.\left(-2x_2^\theta+2x_2^{11}x_2^2+\frac{2}{3}x_2^1(x_2^2)^2
  \right)x_3^1-x_2^1\left(-2x_3^\theta+2x_3^{11}x_3^2+\frac{2}{3}x_3^1
  (x_3^2)^2\right)\right\} 
\label{so51}
\eea
including the "permuted" ones
\bea
 R^{\La^1,0,\La^1}(x_1,x_2,x_3)&=&R^{\La^1,\La^1,0}(x_1,x_3,x_2)\nn
 R^{0,\La^1,\La^1}(x_1,x_2,x_3)&=&R^{\La^1,\La^1,0}(x_2,x_3,x_1)\nn
 R^{\La^2,0,\La^2}(x_1,x_2,x_3)&=&R^{\La^2,\La^2,0}(x_1,x_3,x_2)\nn
 R^{0,\La^2,\La^2}(x_1,x_2,x_3)&=&R^{\La^2,\La^2,0}(x_2,x_3,x_1)\nn
 R^{\La^2,\La^1,\La^2}(x_1,x_2,x_3)&=&R^{\La^1,\La^2,\La^2}(x_2,x_1,x_3)\nn
 R^{\La^2,\La^2,\La^1}(x_1,x_2,x_3)&=&R^{\La^1,\La^2,\La^2}(x_3,x_2,x_1)\nn
 R^{\La^1,2\La^2,\La^1}(x_1,x_2,x_3)&=&R^{2\La^2,\La^1,\La^1}(x_2,x_1,x_3)\nn
 R^{\La^1,\La^1,2\La^2}(x_1,x_2,x_3)&=&R^{2\La^2,\La^1,\La^1}(x_3,x_2,x_1)
\label{so52}
\eea
In (\ref{so51}) the order of the arguments are $R(x_1,x_2,x_3)$. 
In terms of these elementary polynomials the 9 syzygies become
\bea
 0&=&s_1R^{0,\La^2,\La^2}(x_1,x_2,x_3)R^{2\La^2,\La^1,\La^1}(x_1,x_2,x_3)\nn
 &+&t_1R^{\La^2,\La^1,\La^2}(x_1,x_2,x_3)
   R^{\La^2,\La^2,\La^1}(x_1,x_2,x_3)\nn
 &+&u_1R^{0,\La^1,\La^1}(x_1,x_2,x_3)R^{\La^2,\La^2,0}(x_1,x_2,x_3)
  R^{\La^2,0,\La^2}(x_1,x_2,x_3)\nn
 0&=&s_2R^{\La^2,0,\La^2}(x_1,x_2,x_3)R^{\La^1,2\La^2,\La^1}(x_1,x_2,x_3)\nn
 &+&t_2R^{\La^1,\La^2,\La^2}(x_1,x_2,x_3)
   R^{\La^2,\La^2,\La^1}(x_1,x_2,x_3)\nn
 &+&u_2R^{\La^1,0,\La^1}(x_1,x_2,x_3)R^{\La^2,\La^2,0}(x_1,x_2,x_3)
  R^{0,\La^2,\La^2}(x_1,x_2,x_3)\nn
 0&=&s_3R^{\La^2,\La^2,0}(x_1,x_2,x_3)R^{\La^1,\La^1,2\La^2}(x_1,x_2,x_3)\nn
 &+&t_3R^{\La^1,\La^2,\La^2}(x_1,x_2,x_3)
   R^{\La^2,\La^1,\La^2}(x_1,x_2,x_3)\nn
 &+&u_3R^{\La^1,\La^1,0}(x_1,x_2,x_3)R^{\La^2,0,\La^2}(x_1,x_2,x_3)
  R^{0,\La^2,\La^2}(x_1,x_2,x_3)\nn
 0&=&s_4R^{\La^1,\La^2,\La^2}(x_1,x_2,x_3)R^{2\La^2,\La^1,\La^1}
  (x_1,x_2,x_3)\nn
 &+&t_4R^{\La^1,0,\La^1}(x_1,x_2,x_3)
  R^{\La^2,\La^2,0}(x_1,x_2,x_3)R^{\La^2,\La^1,\La^2}(x_1,x_2,x_3)\nn
 &+&u_4R^{\La^1,\La^1,0}(x_1,x_2,x_3)
   R^{\La^2,0,\La^2}(x_1,x_2,x_3)R^{\La^2,\La^2,\La^1}(x_1,x_2,x_3)\nn
 0&=&s_5R^{\La^2,\La^1,\La^2}(x_1,x_2,x_3)R^{\La^1,2\La^2,\La^1}
  (x_1,x_2,x_3)\nn
 &+&t_5R^{0,\La^1,\La^1}(x_1,x_2,x_3)
  R^{\La^2,\La^2,0}(x_1,x_2,x_3)R^{\La^1,\La^2,\La^2}(x_1,x_2,x_3)\nn
 &+&u_5R^{\La^1,\La^1,0}(x_1,x_2,x_3)
   R^{0,\La^2,\La^2}(x_1,x_2,x_3)R^{\La^2,\La^2,\La^1}(x_1,x_2,x_3)\nn
 0&=&s_6R^{\La^2,\La^2,\La^1}(x_1,x_2,x_3)R^{\La^1,\La^1,2\La^2}
  (x_1,x_2,x_3)\nn
 &+&t_6R^{0,\La^1,\La^1}(x_1,x_2,x_3)
  R^{\La^2,0,\La^2}(x_1,x_2,x_3)R^{\La^1,\La^2,\La^2}(x_1,x_2,x_3)\nn
 &+&u_6R^{\La^1,0,\La^1}(x_1,x_2,x_3)
   R^{0,\La^2,\La^2}(x_1,x_2,x_3)R^{\La^2,\La^1,\La^2}(x_1,x_2,x_3)\nn
 0&=&s_7R^{2\La^2,\La^1,\La^1}(x_1,x_2,x_3)R^{\La^1,2\La^2,\La^1}
  (x_1,x_2,x_3)\nn
 &+&t_7R^{\La^1,\La^1,0}(x_1,x_2,x_3)
  R^{\La^2,\La^2,\La^1}(x_1,x_2,x_3)R^{\La^2,\La^2,\La^1}(x_1,x_2,x_3)\nn
 &+&u_7R^{\La^1,0,\La^1}(x_1,x_2,x_3)
   R^{0,\La^1,\La^1}(x_1,x_2,x_3)R^{\La^2,\La^2,0}(x_1,x_2,x_3)
   R^{\La^2,\La^2,0}(x_1,x_2,x_3)\nn
 0&=&s_8R^{2\La^2,\La^1,\La^1}(x_1,x_2,x_3)R^{\La^1,\La^1,2\La^2}
  (x_1,x_2,x_3)\nn
 &+&t_8R^{\La^1,0,\La^1}(x_1,x_2,x_3)
  R^{\La^2,\La^1,\La^2}(x_1,x_2,x_3)R^{\La^2,\La^1,\La^2}(x_1,x_2,x_3)\nn
 &+&u_8R^{\La^1,\La^1,0}(x_1,x_2,x_3)
   R^{0,\La^1,\La^1}(x_1,x_2,x_3)R^{\La^2,0,\La^2}(x_1,x_2,x_3)
   R^{\La^2,0,\La^2}(x_1,x_2,x_3)\nn
 0&=&s_9R^{\La^1,2\La^2,\La^1}(x_1,x_2,x_3)R^{\La^1,\La^1,2\La^2}
  (x_1,x_2,x_3)\nn
 &+&t_9R^{0,\La^1,\La^1}(x_1,x_2,x_3)
  R^{\La^1,\La^2,\La^2}(x_1,x_2,x_3)R^{\La^1,\La^2,\La^2}(x_1,x_2,x_3)\nn
 &+&u_9R^{\La^1,\La^1,0}(x_1,x_2,x_3)
   R^{\La^1,0,\La^1}(x_1,x_2,x_3)R^{0,\La^2,\La^2}(x_1,x_2,x_3)
   R^{0,\La^2,\La^2}(x_1,x_2,x_3)
\eea
where the first 3 correspond to the coupling $(2\La^2,\La^1+\La^2,\La^1+\La^2)$
and permutations thereof, the next 3 correspond
to $(\La^1+2\La^2,\La^1+\La^2,\La^1+\La^2)$ and permutations thereof,
while the last 3 correspond to $(\La^1+2\La^2,\La^1+2\La^2,2\La^1)$ 
and permutations thereof. In the normalization chosen, (\ref{so51}) and
(\ref{so52}), the parameters are worked out to be
\ben
 \begin{array}{lll}
 (t_1,u_1)=(2,4)\hspace{2cm}&(t_2,u_2)=(-2,-4)\hspace{2cm}&(t_3,u_3)=(2,-4)\\
 (t_4,u_4)=(-4,4)&(t_5,u_5)=(4,-4)&(t_6,u_6)=(-4,-4)\\
 (t_7,u_7)=(8,16)&(t_8,u_8)=(8,16)&(t_9,u_9)=(-8,-16)
 \end{array}
\label{so5ut}
\een
where we have fixed $s_\ell=1$, $\ell=1,...,9$.

\subsection{Case of $SL(4)$}

We find that the following polynomials constitute the wave function 
equivalences of basis vectors in the 4 highest weight modules
$\La^1$, $\La^2$, $\La^3$ and $\La^1+\La^3$, respectively
\bea
 K_{100}(x)&=&1\nn
 K_{-110}(x)&=&x^1\nn
 K_{0-11}(x)&=&x^{12}+\hf x^1x^2\nn
 K_{00-1}(x)&=&x^\theta+\hf x^{12}x^3+\hf x^1x^{23}+\frac{1}{6}x^1x^2x^3
\eea
\bea
 K_{010}&=&1\nn
 K_{1-11}(x)&=&x^2\nn
 K_{-101}(x)&=&-x^{12}+\hf x^1x^2\nn
 K_{10-1}(x)&=&x^{23}+\hf x^2x^3\nn
 K_{-11-1}(x)&=&-x^\theta+\hf x^1x^{23}-\hf x^{12}x^3+\frac{1}{3}x^1x^2x^3\nn
 K_{0-10}(x)&=&-x^2x^\theta+x^{12}x^{23}+\frac{1}{12}x^1(x^2)^2x^3
\eea
\bea
 K_{001}(x)&=&1\nn
 K_{01-1}(x)&=&x^3\nn
 K_{1-10}(x)&=&-x^{23}+\hf x^2x^3\nn
 K_{-100}(x)&=&x^\theta-\hf x^1x^{23}-\hf x^{12}x^3+\frac{1}{6}x^1x^2x^3
\eea
\bea
 K_{101}(x)&=&1\nn
 K_{-111}(x)&=&x^1\nn
 K_{11-1}(x)&=&x^3\nn
 K_{0-12}(x)&=&x^{12}+\hf x^1x^2\nn
 K_{2-10}(x)&=&-x^{23}+\hf x^2x^3\nn
 K_{-12-1}(x)&=&x^1x^3\nn
 L_1(x)&=&x^\theta-\frac{3}{2}x^1x^{23}-\hf x^{12}x^3+\frac{2}{3}x^1x^2x^3\nn
 L_2(x)&=&-x^1x^{23}+x^{12}x^3+x^1x^2x^3\nn
 L_3(x)&=&x^\theta+\frac{3}{2}x^{12}x^3+\hf x^1x^{23}+\frac{2}{3}x^1x^2x^3\nn
 K_{-210}(x)&=&2x^1x^\theta-(x^1)^2x^{23}-x^1x^{12}x^3
  +\frac{1}{3}(x^1)^2x^2x^3\nn
 K_{01-2}(x)&=&2x^3x^\theta+x^{12}(x^3)^2+x^1x^{23}x^3+\frac{1}{3}x^1
  x^2(x^3)^2\nn
 K_{1-21}(x)&=&-2x^{12}x^{23}+x^{12}x^2x^3-x^1x^2x^{23}+\hf x^1(x^2)^2x^3\nn
 K_{1-1-1}(x)&=&-2x^{23}x^\theta+x^2x^3x^\theta-x^{12}x^{23}x^3
  -x^1(x^{23})^2+\hf x^{12}x^2(x^3)^2\nn
 &+&\frac{1}{6}x^1x^2x^{23}x^3+\frac{1}{6}x^1(x^2x^3)^2\nn
 K_{-1-11}(x)&=&2x^{12}x^\theta+x^1x^2x^\theta-x^1x^{12}x^{23}-(x^{12})^2x^3
  -\hf (x^1)^2x^2x^{23}\nn
 &-&\frac{1}{6}x^1x^{12}x^2x^3+\frac{1}{6}(x^1x^2)^2x^3\nn
 K_{-10-1}(x)&=&2(x^\theta)^2+\frac{2}{3}x^1x^2x^3x^\theta-\hf(x^1x^{23})^2
  -\hf(x^{12}x^3)^2\nn
 &-&x^1x^{12}x^{23}x^3+\frac{1}{18}(x^1x^2x^3)^2
\eea
The indices denote the weights of the vectors where we have used the obvious 
abbreviation leaving out commas. The special notation 
$L_1,L_2,L_3$ is introduced to characterize the polynomials associated
to the weight 0 which appears with multiplicity 3 in the highest
weight module $\La^1+\La^3$.  
In terms of these, the 18 elementary polynomials are worked out to be
\bea
 R^{\La^1,\La^3,0}(x_1,x_2,x_3)&=&K_{100}(x_1)K_{-100}(x_2)
  -K_{-110}(x_1)K_{1-10}(x_2)\nn
 &+&K_{0-11}(x_1)K_{01-1}(x_2)-K_{00-1}(x_1)K_{001}(x_2)\nn
 R^{\La^1,0,\La^3}(x_1,x_2,x_3)&=&R^{\La^1,\La^3,0}(x_1,x_3,x_2)\nn
 R^{0,\La^1,\La^3}(x_1,x_2,x_3)&=&R^{\La^1,\La^3,0}(x_2,x_3,x_1)\nn
 R^{\La^3,\La^1,0}(x_1,x_2,x_3)&=&R^{\La^1,\La^3,0}(x_2,x_1,x_3)\nn
 R^{\La^3,0,\La^1}(x_1,x_2,x_3)&=&R^{\La^1,\La^3,0}(x_3,x_1,x_2)\nn
 R^{0,\La^3,\La^1}(x_1,x_2,x_3)&=&R^{\La^1,\La^3,0}(x_3,x_2,x_1)\nn
 R^{\La^2,\La^2,0}(x_1,x_2,x_3)&=&K_{010}(x_1)K_{0-10}(x_2)
  -K_{1-11}(x_1)K_{-11-1}(x_2)\nn
 &+&K_{-101}(x_1)K_{10-1}(x_2)+K_{10-1}(x_1)K_{-101}(x_2)\nn
 &-&K_{-11-1}(x_1)K_{1-11}(x_2)+K_{0-10}(x_1)K_{010}(x_2)\nn
 R^{0,\La^2,\La^2}(x_1,x_2,x_3)&=&R^{\La^2,\La^2,0}(x_3,x_2,x_1)\nn
 R^{\La^2,0,\La^2}(x_1,x_2,x_3)&=&R^{\La^2,\La^2,0}(x_1,x_3,x_2)\nn
 R^{\La^2,\La^3,\La^3}(x_1,x_2,x_3)&=&K_{010}(x_1)\left\{K_{1-10}(x_2)
  K_{-100}(x_3)-K_{-100}(x_2)K_{1-10}(x_3)\right\}\nn
 &+&K_{1-11}(x_1)\left\{K_{-100}(x_2)
  K_{01-1}(x_3)-K_{01-1}(x_2)K_{-100}(x_3)\right\}\nn
 &+&K_{-101}(x_1)\left\{K_{01-1}(x_2)K_{1-10}(x_3)
  -K_{1-10}(x_2)K_{01-1}(x_3)\right\}\nn
 &+&K_{10-1}(x_1)\left\{K_{001}(x_2)K_{-100}(x_3)
  -K_{-100}(x_2)K_{001}(x_3)\right\}\nn
 &+&K_{-11-1}(x_1)\left\{K_{1-10}(x_2)K_{001}(x_3)
  -K_{001}(x_2)K_{1-10}(x_3)\right\}\nn
 &+&K_{0-10}(x_1)\left\{K_{001}(x_2)K_{01-1}(x_3)
  -K_{01-1}(x_2)K_{001}(x_3)\right\}\nn
 R^{\La^3,\La^2,\La^3}(x_1,x_2,x_3)&=&R^{\La^2,\La^3,\La^3}(x_2,x_1,x_3)\nn
 R^{\La^3,\La^3,\La^2}(x_1,x_2,x_3)&=&R^{\La^2,\La^3,\La^3}(x_3,x_2,x_1)\nn
 R^{\La^2,\La^1,\La^1}(x_1,x_2,x_3)&=&
  K_{010}(x_1)\left\{K_{0-11}(x_2)K_{00-1}(x_3)
  -K_{00-1}(x_2)K_{0-11}(x_3)\right\}\nn
 &+&K_{1-11}(x_1)\left\{K_{00-1}(x_2)K_{-110}(x_3)
  -K_{-110}(x_2)K_{00-1}(x_3)\right\}\nn
 &+&K_{-101}(x_1)\left\{K_{100}(x_2)K_{00-1}(x_3)
  -K_{00-1}(x_2)K_{100}(x_3)\right\}\nn
 &+&K_{10-1}(x_1)\left\{K_{-110}(x_2)K_{0-11}(x_3)
  -K_{0-11}(x_2)K_{-110}(x_3)\right\}\nn
 &+&K_{-11-1}(x_1)\left\{K_{0-11}(x_2)K_{100}(x_3)
  -K_{100}(x_2)K_{0-11}(x_3)\right\}\nn
 &+&K_{0-10}(x_1)\left\{K_{100}(x_2)K_{-110}(x_3)
  -K_{-110}(x_2)K_{100}(x_3)\right\}\nn
 R^{\La^1,\La^2,\La^1}(x_1,x_2,x_3)&=&R^{\La^2,\La^1,\La^1}(x_2,x_1,x_3)\nn
 R^{\La^1,\La^1,\La^2}(x_1,x_2,x_3)&=&R^{\La^2,\La^1,\La^1}(x_3,x_2,x_1)\nn
 R^{\La^1+\La^3,\La^2,\La^2}(x_1,x_2,x_3)&=&
  2K_{101}(x_1)\left\{K_{-11-1}(x_2)K_{0-10}(x_3)
  -K_{0-10}(x_2)K_{-11-1}(x_3)\right\}\nn
 &+&2K_{-111}(x_1)\left\{K_{0-10}(x_2)K_{10-1}(x_3)
  -K_{10-1}(x_2)K_{0-10}(x_3)\right\}\nn
 &+&2K_{11-1}(x_1)\left\{K_{0-10}(x_2)K_{-101}(x_3)
  -K_{-101}(x_2)K_{0-10}(x_3)\right\}\nn
 &+&2K_{0-12}(x_1)\left\{K_{10-1}(x_2)K_{-11-1}(x_3)
  -K_{-11-1}(x_2)K_{10-1}(x_3)\right\}\nn
 &+&2K_{-12-1}(x_1)\left\{K_{1-11}(x_2)K_{0-10}(x_3)
  -K_{0-10}(x_2)K_{1-11}(x_3)\right\}\nn
 &+&2K_{2-10}(x_1)\left\{K_{-101}(x_2)K_{-11-1}(x_3)
  -K_{-11-1}(x_2)K_{-101}(x_3)\right\}\nn
 &+&K_{01-2}(x_1)\left\{K_{1-11}(x_2)K_{-101}(x_3)
  -K_{-101}(x_2)K_{1-11}(x_3)\right\}\nn
 &+&K_{1-21}(x_1)\left\{K_{010}(x_2)K_{-11-1}(x_3)
  -K_{-11-1}(x_2)K_{010}(x_3)\right\}\nn
 &+&K_{-210}(x_1)\left\{K_{1-11}(x_2)K_{10-1}(x_3)
  -K_{10-1}(x_2)K_{1-11}(x_3)\right\}\nn
 &+&K_{1-1-1}(x_1)\left\{K_{-101}(x_2)K_{010}(x_3)
  -K_{010}(x_2)K_{-101}(x_3)\right\}\nn
 &+&K_{-1-11}(x_1)\left\{K_{10-1}(x_2)K_{010}(x_3)
  -K_{010}(x_2)K_{10-1}(x_3)\right\}\nn
 &+&K_{-10-1}(x_1)\left\{K_{010}(x_2)K_{1-11}(x_3)
  -K_{1-11}(x_2)K_{010}(x_3)\right\}\nn
 &+&L_1(x_1)\left\{K_{010}(x_2)K_{0-10}(x_3)
  -K_{0-10}(x_2)K_{010}(x_3)\right.\nn
 &+&K_{-11-1}(x_2)K_{1-11}(x_3)-K_{1-11}(x_2)K_{-11-1}(x_3)\nn
 &+&\left.K_{10-1}(x_2)K_{-101}(x_3)-K_{-101}(x_2)K_{10-1}(x_3)\right\}\nn
 &+&2L_2(x_1)\left\{K_{0-10}(x_2)K_{010}(x_3)
  -K_{010}(x_2)K_{0-10}(x_3)\right\}\nn
 &+&L_3(x_1)\left\{K_{010}(x_2)K_{0-10}(x_3)
  -K_{0-10}(x_2)K_{010}(x_3)\right.\nn
 &+&K_{-11-1}(x_2)K_{1-11}(x_3)-K_{1-11}(x_2)K_{-11-1}(x_3)\nn
 &+&\left.K_{-101}(x_2)K_{10-1}(x_3)-K_{10-1}(x_2)K_{-101}(x_3)\right\}\nn
 R^{\La^2,\La^1+\La^3,\La^2}(x_1,x_2,x_3)&=&
  R^{\La^1+\La^3,\La^2,\La^2}(x_2,x_1,x_3)\nn
 R^{\La^2,\La^2,\La^1+\La^3}(x_1,x_2,x_3)&=&
  R^{\La^1+\La^3,\La^2,\La^2}(x_3,x_2,x_1)
\label{sl4}
\eea
while the 15 associated syzygies become
\bea
 0&=&s_1R^{\La^1+\La^3,\La^2,\La^2}(x_1,x_2,x_3)R^{\La^2,\La^1+\La^3,\La^2}
  (x_1,x_2,x_3)\nn
 &+&t_1R^{\La^2,\La^2,0}(x_1,x_2,x_3)R^{\La^1,\La^1,\La^2}(x_1,x_2,x_3)
  R^{\La^3,\La^3,\La^2}(x_1,x_2,x_3)\nn
 &+&u_1R^{0,\La^2,\La^2}(x_1,x_2,x_3)R^{\La^2,0,\La^2}(x_1,x_2,x_3)
  R^{\La^3,\La^1,0}(x_1,x_2,x_3)R^{\La^1,\La^3,0}(x_1,x_2,x_3)\nn
 0&=&s_2R^{\La^2,\La^2,\La^1+\La^3}(x_1,x_2,x_3)R^{\La^2,\La^1+\La^3,\La^2}
  (x_1,x_2,x_3)\nn
 &+&t_2R^{0,\La^2,\La^2}(x_1,x_2,x_3)R^{\La^2,\La^1,\La^1}(x_1,x_2,x_3)
  R^{\La^2,\La^3,\La^3}(x_1,x_2,x_3)\nn
 &+&u_2R^{\La^2,\La^2,0}(x_1,x_2,x_3)R^{\La^2,0,\La^2}(x_1,x_2,x_3)
  R^{0,\La^3,\La^1}(x_1,x_2,x_3)R^{0,\La^1,\La^3}(x_1,x_2,x_3)\nn
 0&=&s_3R^{\La^1+\La^3,\La^2,\La^2}(x_1,x_2,x_3)R^{\La^2,\La^2,\La^1+\La^3}
  (x_1,x_2,x_3)\nn
 &+&t_3R^{\La^2,0,\La^2}(x_1,x_2,x_3)R^{\La^1,\La^2,\La^1}(x_1,x_2,x_3)
  R^{\La^3,\La^2,\La^3}(x_1,x_2,x_3)\nn
 &+&u_3R^{0,\La^2,\La^2}(x_1,x_2,x_3)R^{\La^2,\La^2,0}(x_1,x_2,x_3)
  R^{\La^3,0,\La^1}(x_1,x_2,x_3)R^{\La^1,0,\La^3}(x_1,x_2,x_3)\nn
 0&=&s_4R^{\La^1+\La^3,\La^2,\La^2}(x_1,x_2,x_3)R^{\La^2,\La^3,\La^3}
  (x_1,x_2,x_3)\nn
 &+&t_4R^{\La^1,0,\La^3}(x_1,x_2,x_3)R^{\La^2,\La^2,0}(x_1,x_2,x_3)
  R^{\La^3,\La^3,\La^2}(x_1,x_2,x_3)\nn
 &+&u_4R^{\La^1,\La^3,0}(x_1,x_2,x_3)R^{\La^2,0,\La^2}(x_1,x_2,x_3)
  R^{\La^3,\La^2,\La^3}(x_1,x_2,x_3)\nn
 0&=&s_5R^{\La^2,\La^1+\La^3,\La^2}(x_1,x_2,x_3)R^{\La^3,\La^2,\La^3}
  (x_1,x_2,x_3)\nn
 &+&t_5R^{\La^3,\La^1,0}(x_1,x_2,x_3)R^{0,\La^2,\La^2}(x_1,x_2,x_3)
  R^{\La^2,\La^3,\La^3}(x_1,x_2,x_3)\nn
 &+&u_5R^{0,\La^1,\La^3}(x_1,x_2,x_3)R^{\La^2,\La^2,0}(x_1,x_2,x_3)
  R^{\La^3,\La^3,\La^2}(x_1,x_2,x_3)\nn
 0&=&s_6R^{\La^2,\La^2,\La^1+\La^3}(x_1,x_2,x_3)R^{\La^3,\La^3,\La^2}
  (x_1,x_2,x_3)\nn
 &+&t_6R^{0,\La^3,\La^1}(x_1,x_2,x_3)R^{\La^2,0,\La^2}(x_1,x_2,x_3)
  R^{\La^3,\La^2,\La^3}(x_1,x_2,x_3)\nn
 &+&u_6R^{\La^3,0,\La^1}(x_1,x_2,x_3)R^{0,\La^2,\La^2}(x_1,x_2,x_3)
  R^{\La^2,\La^3,\La^3}(x_1,x_2,x_3)\nn
 0&=&s_7R^{\La^2,\La^2,\La^1+\La^3}(x_1,x_2,x_3)R^{\La^1,\La^1,\La^2}
  (x_1,x_2,x_3)\nn
 &+&t_7R^{\La^1,0,\La^3}(x_1,x_2,x_3)R^{0,\La^2,\La^2}(x_1,x_2,x_3)
  R^{\La^2,\La^1,\La^1}(x_1,x_2,x_3)\nn
 &+&u_7R^{0,\La^1,\La^3}(x_1,x_2,x_3)R^{\La^2,0,\La^2}(x_1,x_2,x_3)
  R^{\La^1,\La^2,\La^1}(x_1,x_2,x_3)\nn
 0&=&s_8R^{\La^2,\La^1+\La^3,\La^2}(x_1,x_2,x_3)R^{\La^1,\La^2,\La^1}
  (x_1,x_2,x_3)\nn
 &+&t_8R^{0,\La^3,\La^1}(x_1,x_2,x_3)R^{\La^2,\La^2,0}(x_1,x_2,x_3)
  R^{\La^1,\La^1,\La^2}(x_1,x_2,x_3)\nn
 &+&u_8R^{\La^1,\La^3,0}(x_1,x_2,x_3)R^{0,\La^2,\La^2}(x_1,x_2,x_3)
  R^{\La^2,\La^1,\La^1}(x_1,x_2,x_3)\nn
 0&=&s_9R^{\La^1+\La^3,\La^2,\La^2}(x_1,x_2,x_3)R^{\La^2,\La^1,\La^1}
  (x_1,x_2,x_3)\nn
 &+&t_9R^{\La^3,\La^1,0}(x_1,x_2,x_3)R^{\La^2,0,\La^2}(x_1,x_2,x_3)
  R^{\La^1,\La^2,\La^1}(x_1,x_2,x_3)\nn
 &+&u_9R^{\La^3,0,\La^1}(x_1,x_2,x_3)R^{\La^2,\La^2,0}(x_1,x_2,x_3)
  R^{\La^1,\La^1,\La^2}(x_1,x_2,x_3)\nn
 0&=&s_{10}R^{\La^1+\La^3,\La^2,\La^2}(x_1,x_2,x_3)R^{0,\La^1,\La^3}
  (x_1,x_2,x_3)\nn
 &+&t_{10}R^{\La^1,\La^1,\La^2}(x_1,x_2,x_3)R^{\La^3,\La^2,\La^3}
  (x_1,x_2,x_3)\nn
 &+&u_{10}R^{\La^1,0,\La^3}(x_1,x_2,x_3)R^{\La^3,\La^1,0}(x_1,x_2,x_3)
  R^{0,\La^2,\La^2}(x_1,x_2,x_3)\nn
 0&=&s_{11}R^{\La^2,\La^2,\La^1+\La^3}(x_1,x_2,x_3)R^{\La^1,\La^3,0}
  (x_1,x_2,x_3)\nn
 &+&t_{11}R^{\La^1,\La^2,\La^1}(x_1,x_2,x_3)R^{\La^2,\La^3,\La^3}
  (x_1,x_2,x_3)\nn
 &+&u_{11}R^{\La^1,0,\La^3}(x_1,x_2,x_3)R^{0,\La^3,\La^1}(x_1,x_2,x_3)
  R^{\La^2,\La^2,0}(x_1,x_2,x_3)\nn
 0&=&s_{12}R^{\La^2,\La^1+\La^3,\La^2}(x_1,x_2,x_3)R^{\La^3,0,\La^1}
  (x_1,x_2,x_3)\nn
 &+&t_{12}R^{\La^2,\La^1,\La^1}(x_1,x_2,x_3)R^{\La^3,\La^3,\La^2}
  (x_1,x_2,x_3)\nn
 &+&u_{12}R^{\La^3,\La^1,0}(x_1,x_2,x_3)R^{0,\La^3,\La^1}(x_1,x_2,x_3)
  R^{\La^2,0,\La^2}(x_1,x_2,x_3)\nn
 0&=&s_{13}R^{\La^2,\La^1+\La^3,\La^2}(x_1,x_2,x_3)R^{\La^1,0,\La^3}
  (x_1,x_2,x_3)\nn
 &+&t_{13}R^{\La^1,\La^1,\La^2}(x_1,x_2,x_3)R^{\La^2,\La^3,\La^3}
  (x_1,x_2,x_3)\nn
 &+&u_{13}R^{0,\La^1,\La^3}(x_1,x_2,x_3)R^{\La^1,\La^3,0}(x_1,x_2,x_3)
  R^{\La^2,0,\La^2}(x_1,x_2,x_3)\nn
 0&=&s_{14}R^{\La^2,\La^2,\La^1+\La^3}(x_1,x_2,x_3)R^{\La^3,\La^1,0}
  (x_1,x_2,x_3)\nn
 &+&t_{14}R^{\La^2,\La^1,\La^1}(x_1,x_2,x_3)R^{\La^3,\La^2,\La^3}
  (x_1,x_2,x_3)\nn
 &+&u_{14}R^{0,\La^1,\La^3}(x_1,x_2,x_3)R^{\La^3,0,\La^1}(x_1,x_2,x_3)
  R^{\La^2,\La^2,0}(x_1,x_2,x_3)\nn
 0&=&s_{15}R^{\La^1+\La^3,\La^2,\La^2}(x_1,x_2,x_3)R^{0,\La^3,\La^1}
  (x_1,x_2,x_3)\nn
 &+&t_{15}R^{\La^1,\La^2,\La^1}(x_1,x_2,x_3)R^{\La^3,\La^3,\La^2}
  (x_1,x_2,x_3)\nn
 &+&u_{15}R^{\La^1,\La^3,0}(x_1,x_2,x_3)R^{\La^3,0,\La^1}(x_1,x_2,x_3)
  R^{0,\La^2,\La^2}(x_1,x_2,x_3)
\eea
The first 3 correspond to the coupling $(\La^1+\La^2+\La^3,\La^1+\La^2+\La^3,
2\La^2)$ and permutations thereof, 
the next 3 correspond to the coupling $(\La^1+\La^2+\La^3,\La^2+\La^3,
\La^2+\La^3)$ and permutations thereof,
the next 3 correspond to the coupling $(\La^1+\La^2,\La^1+\La^2,
\La^1+\La^2+\La^3)$ and permutations thereof, whereas the last 6 
correspond to the coupling $(\La^1+\La^3,\La^1+\La^2,\La^2+\La^3)$
and permutations thereof. In the normalization chosen,
(\ref{sl4}), the parameters are worked out to be
\ben
\begin{array}{lll}
 (t_1,u_1)=(4,-4)\hspace{2cm}&(t_2,u_2)=(-4,4)\hspace{2cm}&(t_3,u_3)=(4,-4)\\
 (t_4,u_4)=(2,-2)&(t_5,u_5)=(-2,2)&(t_6,u_6)=(2,2)\\
 (t_7,u_7)=(-2,-2)&(t_8,u_8)=(2,2)&(t_9,u_9)=(2,-2)\\
 (t_{10},u_{10})=(2,-2)&(t_{11},u_{11})=(2,2)&(t_{12},u_{12})=(2,2)\\
 (t_{13},u_{13})=(-2,-2)&(t_{14},u_{14})=(-2,-2)&(t_{15},u_{15})=(-2,2)
\end{array}
\label{sl4ut}
\een
where we have fixed $s_\ell=1$, $\ell=1,...,15$.

\end{document}